\documentclass[runningheads]{llncs}
\usepackage[numbers,sort&compress]{natbib}

\usepackage{amssymb}
\usepackage[hyphens]{url}
\usepackage{hyperref}
\usepackage[hyphenbreaks]{breakurl}

\makeatletter
\@ifundefined{lhs2tex.lhs2tex.sty.read}%
  {\@namedef{lhs2tex.lhs2tex.sty.read}{}%
   \newcommand\SkipToFmtEnd{}%
   \newcommand\EndFmtInput{}%
   \long\def\SkipToFmtEnd#1\EndFmtInput{}%
  }\SkipToFmtEnd

\newcommand\ReadOnlyOnce[1]{\@ifundefined{#1}{\@namedef{#1}{}}\SkipToFmtEnd}
\usepackage{amstext}
\usepackage{amssymb}
\usepackage{stmaryrd}
\DeclareFontFamily{OT1}{cmtex}{}
\DeclareFontShape{OT1}{cmtex}{m}{n}
  {<5><6><7><8>cmtex8
   <9>cmtex9
   <10><10.95><12><14.4><17.28><20.74><24.88>cmtex10}{}
\DeclareFontShape{OT1}{cmtex}{m}{it}
  {<-> ssub * cmtt/m/it}{}

\DeclareFontShape{OT1}{cmtt}{bx}{n}
  {<5><6><7><8>cmtt8
   <9>cmbtt9
   <10><10.95><12><14.4><17.28><20.74><24.88>cmbtt10}{}
\DeclareFontShape{OT1}{cmtex}{bx}{n}
  {<-> ssub * cmtt/bx/n}{}

\newcommand{\anonymous}{\kern0.06em \vbox{\hrule\@width.5em}}

\renewcommand{\leq}{\leqslant}

\usepackage{polytable}

\@ifundefined{mathindent}%
  {\newdimen\mathindent\mathindent\leftmargini}%
  {}%

\def\resethooks{%
  \global\let\SaveRestoreHook\empty
  \global\let\ColumnHook\empty}
\newcommand*{\savecolumns}[1][default]%
  {\g@addto@macro\SaveRestoreHook{\savecolumns[#1]}}
\newcommand*{\restorecolumns}[1][default]%
  {\g@addto@macro\SaveRestoreHook{\restorecolumns[#1]}}
\newcommand*{\aligncolumn}[2]%
  {\g@addto@macro\ColumnHook{\column{#1}{#2}}}

\resethooks

\newcommand{\onelinecommentchars}{\quad-{}- }
\newcommand{\commentbeginchars}{\enskip\{-}
\newcommand{\commentendchars}{-\}\enskip}

\newcommand{\visiblecomments}{%
  \let\onelinecomment=\onelinecommentchars
  \let\commentbegin=\commentbeginchars
  \let\commentend=\commentendchars}

\newcommand{\invisiblecomments}{%
  \let\onelinecomment=\empty
  \let\commentbegin=\empty
  \let\commentend=\empty}

\visiblecomments

\newlength{\blanklineskip}
\newcommand{\hsindent}[1]{\quad}%
\let\hspre\empty
\let\hspost\empty

\EndFmtInput
\makeatother
\ReadOnlyOnce{polycode.fmt}%
\makeatletter

\newcommand{\hsnewpar}[1]%
  {{\parskip=0pt\parindent=0pt\par\vskip #1\noindent}}

\newcommand{\hscodestyle}{}

\newcommand{\sethscode}[1]%
  {\expandafter\let\expandafter\hscode\csname #1\endcsname
   \expandafter\let\expandafter\endhscode\csname end#1\endcsname}

  {\par\noindent
   \advance\leftskip\mathindent
   \hscodestyle
   \let\\=\@normalcr
   \let\hspre\(\let\hspost\)%
   \pboxed}%
  {\endpboxed\)%
   \par\noindent
   \ignorespacesafterend}

  {\hsnewpar\abovedisplayskip
   \advance\leftskip\mathindent
   \hscodestyle
   \let\hspre\(\let\hspost\)%
   \pboxed}%
  {\endpboxed%
   \hsnewpar\belowdisplayskip
   \ignorespacesafterend}

  {\hsnewpar\abovedisplayskip
   \advance\leftskip\mathindent
   \hscodestyle
   \let\\=\@normalcr
   \(\pboxed}%
  {\endpboxed\)%
   \hsnewpar\belowdisplayskip
   \ignorespacesafterend}

\newcommand{\plainhs}{\sethscode{plainhscode}}

\plainhs

  {\hsnewpar\abovedisplayskip
   \advance\leftskip\mathindent
   \hscodestyle
   \let\\=\@normalcr
   \(\parray}%
  {\endparray\)%
   \hsnewpar\belowdisplayskip
   \ignorespacesafterend}

  {\parray}{\endparray}

  {\(\parray}{\endparray\)}

\def\codeframewidth{\arrayrulewidth}
\RequirePackage{calc}

  {\parskip=\abovedisplayskip\par\noindent
   \hscodestyle
   \arrayrulewidth=\codeframewidth
   \tabular{@{}|p{\linewidth-2\arraycolsep-2\arrayrulewidth-2pt}|@{}}%
   \hline\framedhslinecorrect\\{-1.5ex}%
   \let\endoflinesave=\\
   \let\\=\@normalcr
   \(\pboxed}%
  {\endpboxed\)%
   \framedhslinecorrect\endoflinesave{.5ex}\hline
   \endtabular
   \parskip=\belowdisplayskip\par\noindent
   \ignorespacesafterend}

\newcommand{\framedhslinecorrect}[2]%
  {#1[#2]}

  {\(\def\column##1##2{}%
   \let\>\undefined\let\<\undefined\let\\\undefined
   \newcommand\>[1][]{}\newcommand\<[1][]{}\newcommand\\[1][]{}%
   \def\fromto##1##2##3{##3}%
   }{\) }%

  {\let\orighscode=\hscode
   \let\origendhscode=\endhscode
   \def\endhscode{\def\hscode{\endgroup\def\@currenvir{hscode}\\}\begingroup}
   \orighscode\def\hscode{\endgroup\def\@currenvir{hscode}}}%
  {\origendhscode
   \global\let\hscode=\orighscode
   \global\let\endhscode=\origendhscode}%

\makeatother
\EndFmtInput
\usepackage{xcolor}

\definecolor{hsgold2}{RGB/cmyk}{177,149,90/0,0.16,0.49,0.3}
\definecolor{hsgold3}{RGB/cmyk}{190,106,13/0,0.44,0.93,0.25}
\definecolor{hsblue3}{RGB/cmyk}{0,33,132/1,0.65,0,0.35}
\definecolor{hsblue4}{RGB/cmyk}{97,108,132/0.26,0.18,0,0.48}
\definecolor{hsblue5}{RGB/cmyk}{34,50,68/0.5,0.26,0,0.73}
\definecolor{hsred2}{RGB/cmyk}{191,121,103/0,0.4,0.49,0.23}
\definecolor{hsred3}{RGB/cmyk}{171,72,46/0,0.58,0.73,0.33}

\colorlet{HSBLUE3}{hsblue3}

\newcommand*{\mathcolor}{}
\def\mathcolor#1#{\mathcoloraux{#1}}
\newcommand*{\mathcoloraux}[3]{%
  \protect\leavevmode
  \begingroup
    \color#1{#2}#3%
  \endgroup
}

\newcommand{\HSKeyword}[1]{\mathcolor{hsgold3}{\textbf{#1}}}

\newcommand{\HSSpecial}[1]{\mathcolor{hsblue4}{#1}}
\newcommand{\HSSym}[1]{\mathcolor{hsblue4}{\textit{\ensuremath{#1}}}}
\newcommand{\HSCon}[1]{\mathcolor{hsblue3}{\mathit{\ensuremath{#1}}}}
\newcommand{\HSVar}[1]{\mathcolor{hsblue5}{\mathit{\ensuremath{#1}}}}
\newcommand{\HSVarNI}[1]{\mathcolor{hsblue5}{\ensuremath{#1}}}
\newcommand{\HSConNI}[1]{\mathcolor{hsblue3}{\ensuremath{#1}}}

\newcommand{\HT}[1]{\ifdefined\HSCon\HSCon{#1}\else#1\fi}

\newcommand{\HV}[1]{\ifdefined\HSVar\HSVar{#1}\else#1\fi}

\newcommand{\HVNI}[1]{\ifdefined\HSVarNI\HSVarNI{#1}\else#1\fi}
\newcommand{\HTNI}[1]{\ifdefined\HSConNI\HSConNI{#1}\else#1\fi}

\newcommand{\mathnocolor}[2]{#2}
\newcommand{\HSCustomNC}[2]{%
\mathcolor{#1}{\let\mathcolor\mathnocolor%
\ensuremath{#2}}}

\usepackage{tikz}
\usetikzlibrary{positioning}
\usetikzlibrary{calc}
\usetikzlibrary{plotmarks}
\usetikzlibrary{fit}
\usetikzlibrary{shapes}

\usepackage{cleveref}

\renewcommand{\!}{\negthinspace}

\definecolor{C1}{RGB}{0,153,204}
\definecolor{C2}{RGB}{89,0,179}

\newcounter{commentctr}
\renewcommand{\hscodestyle}{\footnotesize}

\makeatletter
  {\hsnewpar\abovedisplayskip
   \hscodestyle
   \let\hspre\(\let\hspost\)%
   \pboxed}
  {\endpboxed%
   \hsnewpar\belowdisplayskip
   \ignorespacesafterend}
  {\vspace*{-5.7em}\hscodestyle
   \let\hspre\(\let\hspost\)%
   \pboxed}
  {\endpboxed\ignorespacesafterend}
\makeatother
\sethscode{myownhscode}

\newenvironment{myhs*}[1][0.95\textwidth]{%
\begin{minipage}{#1}%
}{%
\end{minipage}%
}

\newenvironment{myhs}[1][0.95\textwidth]{%
\begin{myhs*}[#1]
}{%
\end{myhs*}%
}

\newcommand{\librabftorig}{\textsc{LibraBFT}}
\newcommand{\diembft}{\textsc{DiemBFT}}
\newcommand{\librabft}{\textsc{LibraBFT}}
\newcommand{\hotstuff}{\textsc{HotStuff}}

\newcommand{\guydash}{-{\hskip-0.3em}-}

\newcommand{\extendedalt}[2]{\ifx\extendedversion\undefined{#2}\else{#1}\fi}

\newcommand{\extended}[1]{\extendedalt{#1}{}}

\newcommand{\lhsinclude}[1]{}
\newcommand{\extendedversion}{}

\begin{document}

\title{Towards Formal Verification of HotStuff-based Byzantine Fault Tolerant Consensus in Agda\extended{: Extended Version \thanks{This is an extended version of a conference paper with the same title in the proceedings of the 14th NASA Formal Methods Symposium (NFM 2022).  It contains additional details and proof overviews.}}}

\author{
Harold Carr\inst{1}    \and
Christopher Jenkins\inst{2}    \and
Mark Moir\inst{1} \and
Victor Cacciari Miraldo\inst{3} \and
Lisandra Silva\inst{4}}
\authorrunning{H. Carr et al.}
\institute{
Oracle Labs, USA and New Zealand   \and
University of Iowa, USA            \and
Tweag, The Netherlands            \and
Runtime Verification, USA
}

\titlerunning{Towards formal verification of HotStuff-based BFT consensus in Agda}

\date{\today}

\maketitle

\begin{abstract}

\librabft{} is a Byzantine Fault Tolerant (BFT) consensus protocol based on
 \hotstuff{}.  We present an abstract
 model of the protocol underlying \hotstuff{} / \librabft{},
 and formal, machine-checked proofs of their
 core correctness (safety) property and an extended condition that
 enables non-participating parties to verify committed results.
(Liveness properties would be proved for specific implementations, \emph{not} for the
abstract model presented in this paper.)

A key contribution is precisely defining assumptions
about the behavior of honest peers, in an abstract way, \emph{independent} of any particular implementation.  Therefore,
our work is an important step towards proving correctness of an entire class of concrete implementations,
without repeating the hard work of proving correctness of the underlying
protocol. The abstract proofs are for a single configuration (epoch); extending these proofs
across configuration changes is future work.
Our models and proofs are expressed in Agda, and are available in open source.

\end{abstract}

\section{Introduction}
\label{sec:introduction}
There has been
 phenomenal interest in decentralized systems that enable coordination among peers
 that do not necessarily trust each other.  This interest has largely been driven in recent years
by the emergence of blockchain technology.
When the set of participants is limited by \emph{permissioning} or \emph{proof of stake}~\cite{ppcoin-2012,buterin2020combining},
Byzantine Fault Tolerant (BFT)~\cite{10.1145/279227.279229} consensus---which
tolerates some \emph{byzantine} peers actively deviating from the protocol---is of interest.

Due to attractive properties relative to previous BFT consensus protocols,
implementations based on \hotstuff{}~\cite{hotstuff-podc-19} are being
developed and adopted.  For example, the Diem Foundation (formerly Libra Association) was until recently developing \librabft{}
based on \hotstuff{}~\cite{libra-2019-06-28,libra-2020-05-26}.  (\librabft{} was renamed to \diembft{} before being discontinued; other variants are emerging.)

Many published consensus algorithms, including some with manual correctness
 proofs, have been shown to be incorrect~\cite{cachin2017blockchain,redbelly-2019}.  Therefore,
 precise, machine-checked formal verification is essential, particularly for new algorithms
being adopted in practice.
Some of the papers on \hotstuff{} / \librabft{} include brief correctness arguments,
but they lack many details and are not machine checked.  Furthermore,
\librabft{} uses data structures, messages and logic,
that differ significantly from versions on which those informal proofs were based.

Our contributions are as follows:
\begin{itemize}
\item a precise, abstract
 model of the protocol underlying \hotstuff{} / \librabft{};
\item precise formulation of assumptions; and
\item formal, machine-checked proofs of
 core correctness (safety) properties, plus a novel extended condition that
 enables additional functionality.
\end{itemize}

Proving correctness for an \emph{abstraction} of the protocol enables verifying any concrete implementation
by proving that its handlers ensure the assumptions of our abstract proofs.
Our contribution is thus an important step towards proving correctness for an entire class of concrete
 implementations. However, this class does not include all possible variants.  In particular, \diembft{}
recently added an option for committing based on 2-chains, rather than 3-chains, as our work assumes (see~\Cref{sec:abstract:records}).
Adapting our techniques to accommodate 2-chain-based implementations is future work.

This paper focuses on the metatheory around an \emph{abstraction} of a system of peers participating
in the \hotstuff{} / \librabft{} protocol, and assumptions about which peers can participate,
rules that honest peers obey, and the intersection of any two quorums containing at least one honest
peer.  We state and prove key correctness properties, such as that any two committed blocks do not
conflict (i.e., they belong to the same ordered chain of committed blocks).

Our ongoing work~\cite{librabft-agda}
aims to use the results presented in this paper to verify a concrete Haskell
implementation that we have developed based on the Diem Foundation's open-source Rust
implementation~\cite{librabft-rust}.  We have built a system model that can be instantiated with
data types and handlers, yielding a model of a distributed system in which honest peers execute those handlers
and byzantine ones are constrained
only by being unable to forge signatures of honest peers.  We have ported this implementation
to Agda, using a library we have developed~\cite{nasa-submission-syntax-short} to enable the ported code to closely mirror the original,
thus reducing the risk of error.  We have made substantial progress towards proving that the resulting Agda port
satisfies the assumptions established in this paper.

\librabft{} supports configuration changes (also known as epoch changes), whereby
parameters such as the number and identities of participating peers can be changed.  The contribution
described in this paper is an abstract model for a single epoch
and formal, machine-checked proofs of its correctness conditions.  Stating and
proving cross-epoch properties is future work.  Nevertheless, the Haskell implementation we
are verifying supports epoch changes, and our verification infrastructure is prepared
for multiple epochs.  In particular, our abstract modules are parameterized by an ``epoch configuration'' structure.

Our models, definitions and proofs are expressed in Agda~\cite{Norell2008, agda-doc-2.6.1.1}, a dependently-typed
programming language and proof assistant.  We chose Agda for this work because its syntax is similar to Haskell's, making it
easier to develop and have confidence in a model of the implementation we aim to verify.
This paper is intended to be reasonably self contained and
does not require the reader to know Agda.  To that end, we will explain Agda-specific features and syntax
that are important for following the paper.  We encourage interested readers
to explore the open source proofs in detail, and we hope that this paper will provide a
useful overview and guide that will make them more accessible.
For readers who would like to learn about Agda, we recommend starting with
the tutorial in~\cite{plfa}.

In~\Cref{sec:overview}, we
overview salient aspects of \hotstuff{} / \librabft{} to motivate
our approach to abstractly modeling the protocol and formally verifying correctness properties.
In~\Cref{sec:proofs}, we present the definitions used to develop the formal
abstract model of a system of peers participating in the protocol, and to
define traditional and extended correctness properties.  We also
describe their proofs, which are available in open source~\cite{librabft-agda}.  Related work is summarized
in~\Cref{sec:relwork} and concluding remarks and future work appear in~\Cref{sec:conc}.  \extendedalt{}{Additional proof overviews are
included in the extended version of this paper~\cite{librabft-agda-extended}.}

\section{An Overview of \hotstuff{}  / \librabft{}}
\label{sec:overview}

The following overview does \emph{not} fully describe \hotstuff{} and \librabft{}:
it highlights aspects that our abstraction must accommodate to enable our proofs.
Details are in the relevant papers and
repositories~\cite{hotstuff-podc-19,libra-2019-06-28,libra-2020-05-26,librabft-rust,librabft-agda}.

Peers participating in the \hotstuff{} / \librabft{} protocol repeatedly agree to
     extend a chain of \emph{blocks} that is initially empty (represented by a \emph{genesis} record).
Each block identifies (directly or indirectly) the
     block that it extends (or the genesis record if none) via one or more cryptographic
     hashes.  This common \emph{hash chaining}~\cite{Spreitzer1997} technique ensures that
     each block uniquely identifies its predecessor, unless an adversary finds a
     hash collision (e.g., two different blocks that hash to the same value); it is a standard
     assumption that a computationally bounded adversary cannot do so~\citep[Chapter 9]{10.5555/548089}.

We require that two (\emph{honest}) peers that faithfully follow the protocol cannot be convinced to extend the
     chain in conflicting ways: if honest peer \ensuremath{\HSVar{\mathit{p}_{1}}} (resp., \ensuremath{\HSVar{\mathit{p}_{2}}}) determines that block \ensuremath{\HSVar{\mathit{b}_{1}}} (resp. \ensuremath{\HSVar{\mathit{b}_{2}}}) is in the
     chain, then the chain up to one of the blocks \emph{extends} the chain up to the other.
     This must hold even if some (\emph{byzantine}) peers (up to some threshold, as discussed below)
     actively misbehave.

A peer can \emph{propose} to add a new block to a chain, and others can \emph{vote}
   to support the proposal.  A proposed block can include a special \emph{reconfiguration} (\emph{epoch change}) transaction, which would change
the set of peers participating and/or other parameters.
To prevent impersonation,
   messages are signed.

A valid proposal contains or identifies a \emph{quorum certificate} that
represents a \emph{quorum} of votes supporting the previous block.
Based on
   assumptions discussed below, we can be sure that
   any two quorums each contain a vote from at least one honest
   peer in common.
An honest participant
will refuse to vote for a proposal if the requirements for the quorum certificate and
previous blocks are not met.  This ensures that the quorum certificate associated with
each block in a chain satisfies
these requirements, even though some peers that contributed votes to the quorum certificates
may be dishonest.
The conditions for \emph{committing} a block are designed to ensure that honest peers never
contribute votes to two quorums that cause conflicting blocks to be committed.

If a byzantine proposer sends different proposals to different
peers, a quorum of votes for the same proposal may not be generated.
In this case, waiting peers may time out, and initiate a new effort to extend the chain;
this can result in competing proposals to extend the same chain with different blocks.
To distinguish between proposals, each proposed block has an associated \emph{round}, which must be larger than that of
the block that it extends.  Because competing proposals are possible, peers collectively build a \emph{tree} of records, and follow specified rules
to determine when a given proposal has been committed.  The essence of the protocol is in the rules that honest peers must follow, and
what information a peer must verify before committing a proposal.

The goal of this work is
an \emph{abstract} model of the protocol that is independent of all these details, capturing just enough detail
to prove that, if the assumptions are not violated, then honest peers will not commit conflicting proposals.

\section{Correctness properties and proofs}
\label{sec:proofs}

We prove our high-level abstract correctness
properties in module \linebreak \ensuremath{\HSCon{LibraBFT.Abstract.Properties}} (in file \texttt{LibraBFT/Abstract/Properties.agda}), which
receives several module parameters that can be instantiated in order to relate a
particular implementation to the abstract machinery.

\begin{myhs}
\begin{hscode}\SaveRestoreHook
\column{B}{@{}>{\hspre}l<{\hspost}@{}}%
\column{3}{@{}>{\hspre}l<{\hspost}@{}}%
\column{14}{@{}>{\hspre}l<{\hspost}@{}}%
\column{E}{@{}>{\hspre}l<{\hspost}@{}}%
\>[B]{}\HSKeyword{module}\;\HSCon{LibraBFT.Abstract.Properties}{}\<[E]%
\\
\>[B]{}\hsindent{3}{}\<[3]%
\>[3]{}\HSSpecial{(}\HV{\mathcal{E}}\;\mathbin{\HSCon{:}}\;\HSCon{EpochConfig}\HSSpecial{)}\;\HSSpecial{(}\HSCon{UID}\;\mathbin{\HSCon{:}}\;\HSCon{Set}\HSSpecial{)}{}\<[E]%
\\
\>[B]{}\hsindent{3}{}\<[3]%
\>[3]{}\HSSpecial{(}\_{ }{\ensuremath{\stackrel{?}{=}}}UID\!\!\_{ }\;{}\<[14]%
\>[14]{}\mathbin{\HSCon{:}}\;\HSSpecial{(}\HSVar{\mathit{u}_{0}}\;\HSVar{\mathit{u}_{1}}\;\mathbin{\HSCon{:}}\;\HSCon{UID}\HSSpecial{)}\;\mathrel{\HSSym{\to}} \;\HSCon{Dec}\;\HSSpecial{(}\HSVar{\mathit{u}_{0}}\;\mathrel{\HSSym{\equiv}} \;\HSVar{\mathit{u}_{1}}\HSSpecial{)}\HSSpecial{)}{}\<[E]%
\\
\>[B]{}\hsindent{3}{}\<[3]%
\>[3]{}\HSSpecial{(}\HV{\mathcal{V}}\;\mathbin{\HSCon{:}}\;\HSCon{VoteEvidence}\;\HV{\mathcal{E}}\;\HSCon{UID}\HSSpecial{)}{}\<[E]%
\\
\>[B]{}\hsindent{3}{}\<[3]%
\>[3]{}\HSKeyword{where}\;\mathbin{\HSSym{\dots}}{}\<[E]%
\ColumnHook
\end{hscode}\resethooks
\end{myhs}

We first describe \ensuremath{\HSCon{EpochConfig}}; the other module parameters are explained later.
\ensuremath{\HSCon{EpochConfig}}
represents configuration information for an epoch, including:
how many peers participate in the
epoch (\ensuremath{\HSVar{authorsN}}), their identities (\ensuremath{\HSVar{toNodeId}}), and their public keys (\ensuremath{\HSVar{getPubKey}}),
as well as requirements such as each member having a different
public key (\ensuremath{\HV{\hbox{\it PK\guydash{}inj}}}).  Members are identified by values of type \ensuremath{\HSCon{Fin}\;\HSVar{authorsN}}: the
natural numbers less than \ensuremath{\HSVar{authorsN}}; for example, we have
\ensuremath{\HSVar{getPubKey}\;\mathbin{\HSCon{:}}\;\HSCon{Member}\;\mathrel{\HSSym{\to}} \;\HSCon{PK}} where \ensuremath{\HSCon{Member}\;\mathrel{\HSSym{=}}\;\HSCon{Fin}\;\HSVar{authorsN}}.

An \ensuremath{\HSCon{EpochConfig}} also provides \ensuremath{\HSCon{IsQuorum}}, a predicate indicating what
the implementation considers to be a quorum.  The type of \ensuremath{\HSCon{IsQuorum}} is \ensuremath{\HSCon{List}\;\HSCon{Member}\;\mathrel{\HSSym{\to}} \;\HSCon{Set}};
\ensuremath{\HSCon{Set}} is Agda's way of representing an arbitrary type.  This definition is then used to define
another important field of an \ensuremath{\HSCon{EpochConfig}}:

\begin{myhs}
\begin{hscode}\SaveRestoreHook
\column{B}{@{}>{\hspre}l<{\hspost}@{}}%
\column{16}{@{}>{\hspre}l<{\hspost}@{}}%
\column{E}{@{}>{\hspre}l<{\hspost}@{}}%
\>[B]{}\HV{\hbox{\it bft\guydash{}assumption}}\;{}\<[16]%
\>[16]{}\mathbin{\HSCon{:}}\;\HSSym{\forall}\;\HSSpecial{\HSSym{\{\mskip1.5mu} }\HSVar{xs}\;\HSVar{ys}\HSSpecial{\HSSym{\mskip1.5mu\}}}\;\mathrel{\HSSym{\to}} \;\HSCon{IsQuorum}\;\HSVar{xs}\;\mathrel{\HSSym{\to}} \;\HSCon{IsQuorum}\;\HSVar{ys}{}\<[E]%
\\
\>[16]{}\mathrel{\HSSym{\to}} \;\HT{\exists[}{\HSVar{a}}\HT{]}{\HSSpecial{(}\HSVar{a}\;\HT{\in}\;\HSVar{xs}\;\mathbin{\HT{\times}}\;\HSVar{a}\;\HT{\in}\;\HSVar{ys}\;\mathbin{\HT{\times}}\;\HSCon{MetaHonestPK}\;\HSSpecial{(}\HSVar{getPubKey}\;\HSVar{a}\HSSpecial{)}\HSSpecial{)}}{}\<[E]%
\ColumnHook
\end{hscode}\resethooks
\end{myhs}

Here, \ensuremath{\HV{\hbox{\it bft\guydash{}assumption}}} requires that the intersection of any two
quorums contains at least one honest peer.\footnote{\ensuremath{\HSCon{MetaHonestPK}} is a predicate representing whether a peer owning a key behaves honestly.  The \ensuremath{\HSCon{Meta}} prefix identifies this as being part of the formal model and not accessible to implementations, which must not depend on knowing who is honest.
}
Agda supports \emph{implicit} arguments, listed in curly braces, which need not be provided explicitly if
their values can be inferred from context, e.g., \ensuremath{\HSCon{IsQuorum}\;\HSVar{xs}}
implies that \ensuremath{\HSVar{xs}} is of type \ensuremath{\HSCon{List}\;\HSCon{Member}}.  The \ensuremath{\HT{\exists[}{\HSVar{a}}\HT{]}{\cdot }} notation says that there is an \ensuremath{\HSVar{a}}
which satisfies the condition---a product of three conditions, in this case.
The type of \ensuremath{\HSVar{a}} must be implied by context;
here, \ensuremath{\HSVar{a}\;\HT{\in}\;\HSVar{xs}} implies that \ensuremath{\HSVar{a}} is of type \ensuremath{\HSCon{Member}}.

To inherit the correctness properties we prove, an implementation
must provide an \ensuremath{\HSCon{EpochConfig}} as a module parameter.  Part
of constructing it is proving \ensuremath{\HV{\hbox{\it bft\guydash{}assumption}}} based on whatever assumptions and definition of \ensuremath{\HSCon{IsQuorum}} the implementation uses.
One common approach is to assume \ensuremath{\HSVar{n}} peers with equal ``voting power''\negthinspace,
at most \ensuremath{\HSVar{f}} of which are byzantine, and to ensure that $n > 3f$; in this case,
a set of peers is a quorum iff it contains at least $2n / 3$ distinct peers.
\ensuremath{\HSCon{LibraBFT.Abstract.BFT}} contains a lemma that can be used to prove that such
assumptions imply \ensuremath{\HV{\hbox{\it bft\guydash{}assumption}}}.  The lemma is sufficiently general to
accommodate \librabft{}'s approach of
assigning (potentially non-uniform) \emph{voting power} to peers, and considering a set of
peers to be a quorum iff its combined voting power exceeds two thirds of the total voting power.

The remainder of this section is in context of a single \ensuremath{\HSCon{EpochConfig}} called \ensuremath{\HV{\mathcal{E}}}.

\subsection{Abstract \ensuremath{\HSCon{Record}}s and \ensuremath{\HT{RecordChain}}s}
\label{sec:abstract:records}

A \ensuremath{\HSCon{Record}} can be a \ensuremath{\HSCon{Block}}, a quorum certificate (\ensuremath{\HSCon{QC}}) or the epoch's \emph{genesis} (initial)
\ensuremath{\HSCon{Record}}; precise definitions are below.  (These are \emph{abstract} records that may not correlate
closely to data structures and message formats used
by an implementation; for example, in \librabft{}, blocks \emph{contain} the previous QC.)
\hotstuff-based algorithms grow a tree of \ensuremath{\HSCon{Record}}s rooted at the epoch's genesis record, where nodes
contain a \ensuremath{\HSCon{Block}} or a \ensuremath{\HSCon{QC}}.  Paths (called \ensuremath{\HT{RecordChain}}s) from the root
begin with the genesis record and then alternate between \ensuremath{\HSCon{Block}}s and \ensuremath{\HSCon{QC}}s.
For example, the existence of a
path from the root to a record \ensuremath{\HSVar{r}} is captured by the type \ensuremath{\HT{RecordChain}\;\HSVar{r}} being inhabited. \Cref{fig:abstract:blocktree} illustrates
a tree of \ensuremath{\HSCon{Record}}s.

\begin{figure*}
\centering
\resizebox{.8\textwidth}{!}{%
\begin{tikzpicture}[txt/.style={scale=1.6}]
\node [txt] (hinit) {\ensuremath{\HSVar{genesis}}};
\node [txt, right = of hinit] (b0) {\ensuremath{\HSVar{\mathit{b}_{0}}}};
\node [txt, right = of b0]   (q0) {\ensuremath{\HSVar{\mathit{q}_{0}}}};
\node [txt, right = of q0]   (b1) {\ensuremath{\HSVar{\mathit{b}_{1}}}};
\node [txt, right = of b1]   (q1) {\ensuremath{\HSVar{\mathit{q}_{1}}}};
\node [txt, above = of b1]   (b2) {\ensuremath{\HSVar{\mathit{b}_{2}}}};
\node [txt, right = of b2]   (q2) {\ensuremath{\HSVar{\mathit{q}_{2}}}};
\node [txt, right = of q1]   (b3) {\ensuremath{\HSVar{\mathit{b}_{3}}}};
\node [txt, right = of b3]   (q3) {\ensuremath{\HSVar{\mathit{q}_{3}}}};
\node [txt, below = of b3]   (b4) {\ensuremath{\HSVar{\mathit{b}_{4}}}};
\node [txt, right = of b4]   (q4) {\ensuremath{\HSVar{\mathit{q}_{4}}}};
\node [txt, right = of q3]   (b5) {\ensuremath{\HSVar{\mathit{b}_{5}}}};
\node [txt, right = of b5]   (q5) {\ensuremath{\HSVar{\mathit{q}_{5}}}};
\node [txt, right = of q4]   (b6) {\ensuremath{\HSVar{\mathit{b}_{6}}}};
\node [txt, right = of b6]   (q6) {\ensuremath{\HSVar{\mathit{q}_{6}}}};

\draw[->] (q6) -- (b6);
\draw[->] (b6) -- (q4);
\draw[->] (q4) -- (b4);
\draw[->] (b4) to[in=270 , out=180] (q1);
\draw[->] (q5) -- (b5);
\draw[->] (b5) -- (q3);
\draw[->] (q3) -- (b3);
\draw[->] (b3) -- (q1);
\draw[->] (q1) -- (b1);
\draw[->] (b1) -- (q0);
\draw[->] (q0) -- (b0);
\draw[->] (b0) -- (hinit);
\draw[->] (q2) -- (b2);
\draw[->] (b2) to[in=90 , out=180] (q0);

\node (box) [fit=(b0) (q2) (q6) , draw , dashed, rounded corners=2mm] {};
\node at (box.north) [scale = 1.6, above, inner sep=1mm]
  {\textit{Tree of Records}};

\draw [dotted] (hinit.south) -- ($ (q1.south) - (0.5,0) $)
               to[out=270, in=180] (b4.north) -- (q6.north);
\node at (b6.north) [scale = 1.6, above , inner sep=1mm] {\ensuremath{\HT{RecordChain}\;\HSSpecial{(}\HSCon{Q}\;\HSVar{\mathit{q}_{6}}\HSSpecial{)}}};

\end{tikzpicture}}
\caption{A tree of \ensuremath{\HSCon{Record}}s with a \ensuremath{\HT{RecordChain}} from \ensuremath{\HSVar{genesis}} to abstract \ensuremath{\HSCon{Record}} \ensuremath{\HSCon{Q}\;\HSVar{\mathit{q}_{6}}}.}
\label{fig:abstract:blocktree}
\end{figure*}

While typical implementations carry more information, abstractly, a \ensuremath{\HSCon{Block}} comprises
 its round number, an identifier of type \ensuremath{\HSCon{UID}} for itself and for the quorum certificate it extends, if any
(a value of type \ensuremath{\HSCon{Maybe}\;\HSCon{UID}} is either \ensuremath{\HSVar{nothing}} or \ensuremath{\HSVar{just}\;\HSVar{x}} for some \ensuremath{\HSVar{x}} of type \ensuremath{\HSCon{UID}}).
\ensuremath{\HSCon{UID}} can be any type that has decidable equality, as represented by the
second and third module parameters; these are passed
to other modules in the \ensuremath{\HSCon{Abstract}} namespace as needed.
Definitions below are in modules \ensuremath{\HSCon{LibraBFT.Abstract.Records}} and \ensuremath{\HSCon{LibraBFT.Abstract.RecordChain}}.

Typical implementations obtain a \ensuremath{\HSCon{Block}}'s id by applying a cryptographic hash function to
some or all of its contents; thus identifiers may not be unique.  Our correctness properties
are therefore proved modulo ``injectivity failures'' on (supposedly) unique ids.
We do \emph{not} assume that such injectivity failures
do not exist, which
would make our proofs meaningless because they \emph{can} occur in practice, however unlikely.  We elaborate below
and in~\Cref{sec:relwork,sec:conc}.

  Abstractly, a \ensuremath{\HSCon{Vote}} is by a member of the epoch, for a round and \ensuremath{\HSCon{Block}} id.

\begin{minipage}{0.5\textwidth}
\begin{myhs}
\begin{hscode}\SaveRestoreHook
\column{B}{@{}>{\hspre}l<{\hspost}@{}}%
\column{3}{@{}>{\hspre}l<{\hspost}@{}}%
\column{11}{@{}>{\hspre}l<{\hspost}@{}}%
\column{15}{@{}>{\hspre}l<{\hspost}@{}}%
\column{20}{@{}>{\hspre}l<{\hspost}@{}}%
\column{E}{@{}>{\hspre}l<{\hspost}@{}}%
\>[B]{}\HSKeyword{record}\;\HSCon{Block}\;{}\<[15]%
\>[15]{}\mathbin{\HSCon{:}}\;\HSCon{Set}\;\HSKeyword{where}{}\<[E]%
\\
\>[B]{}\hsindent{3}{}\<[3]%
\>[3]{}\HSKeyword{constructor}\;\HSVar{mkBlock}{}\<[E]%
\\
\>[B]{}\hsindent{3}{}\<[3]%
\>[3]{}\HSKeyword{field}\;{}\<[11]%
\>[11]{}\HSVar{bRound}\;{}\<[20]%
\>[20]{}\mathbin{\HSCon{:}}\;\HSCon{Round}{}\<[E]%
\\
\>[11]{}\HSVar{bId}\;{}\<[20]%
\>[20]{}\mathbin{\HSCon{:}}\;\HSCon{UID}{}\<[E]%
\\
\>[11]{}\HSVar{bPrevQC}\;{}\<[20]%
\>[20]{}\mathbin{\HSCon{:}}\;\HSCon{Maybe}\;\HSCon{UID}{}\<[E]%
\ColumnHook
\end{hscode}\resethooks
\end{myhs}
\end{minipage}
\begin{minipage}{0.5\textwidth}
\begin{myhs}
\begin{hscode}\SaveRestoreHook
\column{B}{@{}>{\hspre}l<{\hspost}@{}}%
\column{3}{@{}>{\hspre}l<{\hspost}@{}}%
\column{10}{@{}>{\hspre}l<{\hspost}@{}}%
\column{21}{@{}>{\hspre}l<{\hspost}@{}}%
\column{E}{@{}>{\hspre}l<{\hspost}@{}}%
\>[B]{}\HSKeyword{record}\;\HSCon{Vote}\;\mathbin{\HSCon{:}}\;\HSCon{Set}\;\HSKeyword{where}{}\<[E]%
\\
\>[B]{}\hsindent{3}{}\<[3]%
\>[3]{}\HSKeyword{constructor}\;\HSVar{mkVote}{}\<[E]%
\\
\>[B]{}\hsindent{3}{}\<[3]%
\>[3]{}\HSKeyword{field}\;{}\<[10]%
\>[10]{}\HSVar{vRound}\;{}\<[21]%
\>[21]{}\mathbin{\HSCon{:}}\;\HSCon{Round}{}\<[E]%
\\
\>[10]{}\HSVar{vMember}\;{}\<[21]%
\>[21]{}\mathbin{\HSCon{:}}\;\HSCon{Member}{}\<[E]%
\\
\>[10]{}\HSVar{vBlockUID}\;{}\<[21]%
\>[21]{}\mathbin{\HSCon{:}}\;\HSCon{UID}{}\<[E]%
\ColumnHook
\end{hscode}\resethooks
\end{myhs}
\end{minipage}

  A quorum certificate (\ensuremath{\HSCon{QC}}) represents enough \ensuremath{\HSCon{Vote}}s to \emph{certify}
that a \ensuremath{\HSCon{Block}} has been accepted by a quorum of members.
It includes the \ensuremath{\HSCon{Block}}'s id and round, and
a list of \ensuremath{\HSCon{Vote}}s and evidence that the QC is ``valid'' (representing
properties that honest peers verify before accepting the QC), i.e.,:
\begin{enumerate}
\item The list of voting \ensuremath{\HSCon{Member}}s represents a quorum.
\item All \ensuremath{\HSCon{Vote}}s are for the \ensuremath{\HSCon{Block}}'s id.
\item All \ensuremath{\HSCon{Vote}}s are for the same round.
\end{enumerate}
Honest peers accept a (concrete) \ensuremath{\HSCon{Vote}} only if it satisfies implementation-specific conditions
captured by the module parameter \ensuremath{\HV{\mathcal{V}}} of type
\ensuremath{\HSCon{VoteEvidence}\;\HV{\mathcal{E}}\;\HSCon{UID}}, an implementation-specific predicate on abstract \ensuremath{\HSCon{Vote}}s.
To enable proofs to access the verified conditions, we add a fourth
coherence clause to QCs:

\newcommand\setItemnumber[1]{\setcounter{enumi}{\numexpr#1-1\relax}}
\begin{enumerate}
\setItemnumber{4}
\item For each \ensuremath{\HSCon{Vote}} in the \ensuremath{\HSCon{QC}}, there is evidence that a message
was sent containing a concrete representation of the (abstract) \ensuremath{\HSCon{Vote}} that satisfies
the implementation-specific conditions.
\end{enumerate}

Putting this all together, we have:

\begin{myhs}
\begin{hscode}\SaveRestoreHook
\column{B}{@{}>{\hspre}l<{\hspost}@{}}%
\column{3}{@{}>{\hspre}l<{\hspost}@{}}%
\column{10}{@{}>{\hspre}l<{\hspost}@{}}%
\column{24}{@{}>{\hspre}l<{\hspost}@{}}%
\column{51}{@{}>{\hspre}l<{\hspost}@{}}%
\column{69}{@{}>{\hspre}l<{\hspost}@{}}%
\column{E}{@{}>{\hspre}l<{\hspost}@{}}%
\>[B]{}\HSKeyword{record}\;\HSCon{QC}\;\mathbin{\HSCon{:}}\;\HSCon{Set}\;\HSKeyword{where}{}\<[E]%
\\
\>[B]{}\hsindent{3}{}\<[3]%
\>[3]{}\HSKeyword{constructor}\;\HSVar{mkQC}{}\<[E]%
\\
\>[B]{}\hsindent{3}{}\<[3]%
\>[3]{}\HSKeyword{field}\;{}\<[10]%
\>[10]{}\HSVar{qRound}\;{}\<[24]%
\>[24]{}\mathbin{\HSCon{:}}\;\HSCon{Round}{}\<[E]%
\\
\>[10]{}\HSVar{qCertBlockId}\;{}\<[24]%
\>[24]{}\mathbin{\HSCon{:}}\;\HSCon{UID}{}\<[E]%
\\
\>[10]{}\HV{q\!Votes}\;{}\<[24]%
\>[24]{}\mathbin{\HSCon{:}}\;\HSCon{List}\;\HSCon{Vote}{}\<[E]%
\\
\>[10]{}\HV{\hbox{\it q\!Votes\guydash{}C1}}\;{}\<[24]%
\>[24]{}\mathbin{\HSCon{:}}\;\HSCon{IsQuorum}\;\HSSpecial{(}\HV{\hbox{\it List\guydash{}map}}\;\HSVar{vMember}\;\HV{q\!Votes}\HSSpecial{)}{}\<[E]%
\\
\>[10]{}\HV{\hbox{\it q\!Votes\guydash{}C2}}\;{}\<[24]%
\>[24]{}\mathbin{\HSCon{:}}\;\HSCon{All}\;\HSSpecial{(}\HSSym{\lambda} \;\HSVar{v}\;\mathrel{\HSSym{\to}} \;\HSVar{vBlockUID}\;\HSVar{v}\;{}\<[51]%
\>[51]{}\mathrel{\HSSym{\equiv}} \;\HSVar{qCertBlockId}\HSSpecial{)}\;{}\<[69]%
\>[69]{}\HV{q\!Votes}{}\<[E]%
\\
\>[10]{}\HV{\hbox{\it q\!Votes\guydash{}C3}}\;{}\<[24]%
\>[24]{}\mathbin{\HSCon{:}}\;\HSCon{All}\;\HSSpecial{(}\HSSym{\lambda} \;\HSVar{v}\;\mathrel{\HSSym{\to}} \;\HSVar{vRound}\;\HSVar{v}\;{}\<[51]%
\>[51]{}\mathrel{\HSSym{\equiv}} \;\HSVar{qRound}\HSSpecial{)}\;{}\<[69]%
\>[69]{}\HV{q\!Votes}{}\<[E]%
\\
\>[10]{}\HV{\hbox{\it q\!Votes\guydash{}C4}}\;{}\<[24]%
\>[24]{}\mathbin{\HSCon{:}}\;\HSCon{All}\;\HV{\mathcal{V}}\;\HV{q\!Votes}{}\<[E]%
\ColumnHook
\end{hscode}\resethooks
\end{myhs}

\ensuremath{\HSCon{All}} (from the Agda standard library) accepts a predicate and a list, and
requires that each element of the list satisfies the predicate.

Next, we define a \ensuremath{\HSCon{Record}} to be either a \ensuremath{\HSCon{Block}}, a \ensuremath{\HSCon{QC}}, or the special
genesis record \ensuremath{\HSCon{I}}.  There is a constructor for each case, and the \ensuremath{\HSCon{B}}
and \ensuremath{\HSCon{Q}} constructors take arguments of the appropriate type to form a \ensuremath{\HSCon{Record}}.

\begin{myhs}
\begin{hscode}\SaveRestoreHook
\column{B}{@{}>{\hspre}l<{\hspost}@{}}%
\column{3}{@{}>{\hspre}l<{\hspost}@{}}%
\column{14}{@{}>{\hspre}l<{\hspost}@{}}%
\column{18}{@{}>{\hspre}l<{\hspost}@{}}%
\column{E}{@{}>{\hspre}l<{\hspost}@{}}%
\>[B]{}\HSKeyword{data}\;\HSCon{Record}\;\mathbin{\HSCon{:}}\;\HSCon{Set}\;\HSKeyword{where}{}\<[E]%
\\
\>[B]{}\hsindent{3}{}\<[3]%
\>[3]{}\HSCon{I}\;\mathbin{\HSCon{:}}\;{}\<[18]%
\>[18]{}\HSCon{Record}{}\<[E]%
\\
\>[B]{}\hsindent{3}{}\<[3]%
\>[3]{}\HSCon{B}\;\mathbin{\HSCon{:}}\;\HSCon{Block}\;{}\<[14]%
\>[14]{}\mathrel{\HSSym{\to}} \;{}\<[18]%
\>[18]{}\HSCon{Record}{}\<[E]%
\\
\>[B]{}\hsindent{3}{}\<[3]%
\>[3]{}\HSCon{Q}\;\mathbin{\HSCon{:}}\;\HSCon{QC}\;{}\<[14]%
\>[14]{}\mathrel{\HSSym{\to}} \;{}\<[18]%
\>[18]{}\HSCon{Record}{}\<[E]%
\ColumnHook
\end{hscode}\resethooks
\end{myhs}

  We then say that a record \ensuremath{\HSVar{r'}} \emph{extends} another record \ensuremath{\HSVar{r}},
denoted \ensuremath{\HSVar{r}\;\HT{\leftarrow}\;\HV{r^{\prime}}}, whenever one of the following conditions is met:
\begin{enumerate}
\item \ensuremath{\HSVar{r}} is the genesis \ensuremath{\HSCon{Record}} and \ensuremath{\HSVar{r'}} is a \ensuremath{\HSCon{Block}} for round greater than 0 and not identifying any previous \ensuremath{\HSCon{Block}}.
\item \ensuremath{\HSVar{r}} is a \ensuremath{\HSCon{QC}} and \ensuremath{\HSVar{r'}} is a \ensuremath{\HSCon{Block}} with a round higher than \ensuremath{\HSVar{r}}'s and with a \ensuremath{\HSVar{bPrevQC}} field identifying \ensuremath{\HSVar{r}}.
\item \ensuremath{\HSVar{r}} is a \ensuremath{\HSCon{Block}} and \ensuremath{\HSVar{r'}} is a \ensuremath{\HSCon{QC}} certifying \ensuremath{\HSVar{r}}.
\end{enumerate}
We capture these conditions in the following Agda datatype; \ensuremath{\HT{\_\!\!\ensuremath{\leftarrow}\!\!\_}} indicates
that \ensuremath{\HT{\leftarrow}} is an infix operator with two arguments.  Values of this type can be constructed
using one of three constructors (\ensuremath{\HT{I\!\!\leftarrow\!\!B}}, \ensuremath{\HT{Q\!\!\leftarrow\!\!B}} or \ensuremath{\HT{B\!\!\leftarrow\!\!Q}}), each of which requires several arguments
to establish a value of \ensuremath{\HT{\_\!\!\ensuremath{\leftarrow}\!\!\_}} for a pair of \ensuremath{\HSCon{Record}}s.

\begin{myhs}
\begin{hscode}\SaveRestoreHook
\column{B}{@{}>{\hspre}l<{\hspost}@{}}%
\column{4}{@{}>{\hspre}l<{\hspost}@{}}%
\column{6}{@{}>{\hspre}l<{\hspost}@{}}%
\column{13}{@{}>{\hspre}l<{\hspost}@{}}%
\column{E}{@{}>{\hspre}l<{\hspost}@{}}%
\>[4]{}\HSKeyword{data}\;\HT{\_\!\!\ensuremath{\leftarrow}\!\!\_}\;\mathbin{\HSCon{:}}\;\HSCon{Record}\;\mathrel{\HSSym{\to}} \;\HSCon{Record}\;\mathrel{\HSSym{\to}} \;\HSCon{Set}\;\HSKeyword{where}{}\<[E]%
\\
\>[4]{}\hsindent{2}{}\<[6]%
\>[6]{}\HT{I\!\!\leftarrow\!\!B}\;{}\<[13]%
\>[13]{}\mathbin{\HSCon{:}}\;\HSSym{\forall}\;\HSSpecial{\HSSym{\{\mskip1.5mu} }\HSVar{b}\HSSpecial{\HSSym{\mskip1.5mu\}}}\;\mathrel{\HSSym{\to}} \;\HSVar{0}\;\HSVar{<}\;\HSVar{getRound}\;\HSVar{b}\;\mathrel{\HSSym{\to}} \;\HSVar{bPrevQC}\;\HSVar{b}\;\mathrel{\HSSym{\equiv}} \;\HSVar{nothing}{}\<[E]%
\\
\>[13]{}\mathrel{\HSSym{\to}} \;\HSCon{I}\;\HT{\leftarrow}\;\HSSpecial{(}\HSCon{B}\;\HSVar{b}\HSSpecial{)}{}\<[E]%
\\
\>[4]{}\hsindent{2}{}\<[6]%
\>[6]{}\HT{Q\!\!\leftarrow\!\!B}\;{}\<[13]%
\>[13]{}\mathbin{\HSCon{:}}\;\HSSym{\forall}\;\HSSpecial{\HSSym{\{\mskip1.5mu} }\HSVar{q}\;\HSVar{b}\HSSpecial{\HSSym{\mskip1.5mu\}}}\;\mathrel{\HSSym{\to}} \;\HSVar{getRound}\;\HSVar{q}\;\HSVar{<}\;\HSVar{getRound}\;\HSVar{b}{}\<[E]%
\\
\>[13]{}\mathrel{\HSSym{\to}} \;\HSVar{just}\;\HSSpecial{(}\HSVar{qCertBlockId}\;\HSVar{q}\HSSpecial{)}\;\mathrel{\HSSym{\equiv}} \;\HSVar{bPrevQC}\;\HSVar{b}{}\<[E]%
\\
\>[13]{}\mathrel{\HSSym{\to}} \;\HSCon{Q}\;\HSVar{q}\;\HT{\leftarrow}\;\HSCon{B}\;\HSVar{b}{}\<[E]%
\\
\>[4]{}\hsindent{2}{}\<[6]%
\>[6]{}\HT{B\!\!\leftarrow\!\!Q}\;{}\<[13]%
\>[13]{}\mathbin{\HSCon{:}}\;\HSSym{\forall}\;\HSSpecial{\HSSym{\{\mskip1.5mu} }\HSVar{b}\;\HSVar{q}\HSSpecial{\HSSym{\mskip1.5mu\}}}\;\mathrel{\HSSym{\to}} \;\HSVar{getRound}\;\HSVar{q}\;\mathrel{\HSSym{\equiv}} \;\HSVar{getRound}\;\HSVar{b}\;\mathrel{\HSSym{\to}} \;\HSVar{bId}\;\HSVar{b}\;\mathrel{\HSSym{\equiv}} \;\HSVar{qCertBlockId}\;\HSVar{q}{}\<[E]%
\\
\>[13]{}\mathrel{\HSSym{\to}} \;\HSCon{B}\;\HSVar{b}\;\HT{\leftarrow}\;\HSCon{Q}\;\HSVar{q}{}\<[E]%
\ColumnHook
\end{hscode}\resethooks
\end{myhs}

\ensuremath{\HT{RecordChain}}s are in the reflexive,
transitive closure of \ensuremath{\HT{\_\!\!\ensuremath{\leftarrow}\!\!\_}}, starting at the genesis record \ensuremath{\HSCon{I}}. Sometimes, we reason
about paths starting at records other than \ensuremath{\HSCon{I}}; we therefore define
\ensuremath{\HT{RecordChain}} using the more specific \ensuremath{\HT{RecordChainFrom}}.

\begin{myhs}
\begin{hscode}\SaveRestoreHook
\column{B}{@{}>{\hspre}l<{\hspost}@{}}%
\column{4}{@{}>{\hspre}l<{\hspost}@{}}%
\column{11}{@{}>{\hspre}l<{\hspost}@{}}%
\column{E}{@{}>{\hspre}l<{\hspost}@{}}%
\>[B]{}\HSKeyword{data}\;\HT{RecordChainFrom}\;\HSSpecial{(}\HSVar{o}\;\mathbin{\HSCon{:}}\;\HSCon{Record}\HSSpecial{)}\;\mathbin{\HSCon{:}}\;\HSCon{Record}\;\mathrel{\HSSym{\to}} \;\HSCon{Set}\;\HSKeyword{where}{}\<[E]%
\\
\>[B]{}\hsindent{4}{}\<[4]%
\>[4]{}\HSVar{empty}\;{}\<[11]%
\>[11]{}\mathbin{\HSCon{:}}\;\HT{RecordChainFrom}\;\HSVar{o}\;\HSVar{o}{}\<[E]%
\\
\>[B]{}\hsindent{4}{}\<[4]%
\>[4]{}\HSVar{step}\;{}\<[11]%
\>[11]{}\mathbin{\HSCon{:}}\;\HSSym{\forall}\;\HSSpecial{\HSSym{\{\mskip1.5mu} }\HSVar{r}\;\HSVar{r'}\HSSpecial{\HSSym{\mskip1.5mu\}}}\;\mathrel{\HSSym{\to}} \;\HT{RecordChainFrom}\;\HSVar{o}\;\HSVar{r}{}\<[E]%
\\
\>[11]{}\mathrel{\HSSym{\to}} \;\HSVar{r}\;\HT{\leftarrow}\;\HSVar{r'}{}\<[E]%
\\
\>[11]{}\mathrel{\HSSym{\to}} \;\HT{RecordChainFrom}\;\HSVar{o}\;\HSVar{r'}{}\<[E]%
\\[\blanklineskip]%
\>[B]{}\HT{RecordChain}\;\mathbin{\HSCon{:}}\;\HSCon{Record}\;\mathrel{\HSSym{\to}} \;\HSCon{Set}{}\<[E]%
\\
\>[B]{}\HT{RecordChain}\;\mathrel{\HSSym{=}}\;\HT{RecordChainFrom}\;\HSCon{I}{}\<[E]%
\ColumnHook
\end{hscode}\resethooks
\end{myhs}

Next, we present definitions needed to specify when a \ensuremath{\HSCon{Block}} can be
committed.  For $k > 0$, a \ensuremath{\HT{\mathbb{K}\hbox{\it \guydash{}chain}}} is a sequence of $k$ \ensuremath{\HSCon{Block}}s, each of which is extended by a \ensuremath{\HSCon{QC}},
such that each \ensuremath{\HSCon{Block}} (except the first) extends the \ensuremath{\HSCon{QC}} that extends the previous \ensuremath{\HSCon{Block}}.  Furthermore,
each adjacent pair of \ensuremath{\HSCon{Block}}s must satisfy the relation \ensuremath{\HSCon{R}}, which can be instantiated with \ensuremath{\HSCon{Simple}}
(which holds for any pair of \ensuremath{\HSCon{Block}}s) or \ensuremath{\HSCon{Contig}} (which holds only if the rounds of the two \ensuremath{\HSCon{Block}}s are contiguous: the second \ensuremath{\HSCon{Block}}'s round
is one greater than that of the first; the first parameter to \ensuremath{\HSCon{R}} enables a definition of \ensuremath{\HSCon{Contig}}
that does not require a predecessor for the first \ensuremath{\HSCon{Block}}; see module
\ensuremath{\HSCon{LibraBFT.Abstract.RecordChain}}).  \ensuremath{\HT{\mathbb{K}\hbox{\it \guydash{}chain}}}s are
defined as follows.

\begin{myhs}
\begin{hscode}\SaveRestoreHook
\column{B}{@{}>{\hspre}l<{\hspost}@{}}%
\column{4}{@{}>{\hspre}l<{\hspost}@{}}%
\column{6}{@{}>{\hspre}l<{\hspost}@{}}%
\column{15}{@{}>{\hspre}l<{\hspost}@{}}%
\column{31}{@{}>{\hspre}l<{\hspost}@{}}%
\column{E}{@{}>{\hspre}l<{\hspost}@{}}%
\>[B]{}\HSKeyword{data}\;\HT{\mathbb{K}\hbox{\it \guydash{}chain}}\;\HSSpecial{(}\HSCon{R}\;\mathbin{\HSCon{:}}\;\HT{\mathbb{N}}\;\mathrel{\HSSym{\to}} \;\HSCon{Record}\;\mathrel{\HSSym{\to}} \;\HSCon{Record}\;\mathrel{\HSSym{\to}} \;\HSCon{Set}\HSSpecial{)}{}\<[E]%
\\
\>[B]{}\hsindent{6}{}\<[6]%
\>[6]{}\mathbin{\HSCon{:}}\;\HSSpecial{(}\HSVar{k}\;\mathbin{\HSCon{:}}\;\HT{\mathbb{N}}\HSSpecial{)}\;\HSSpecial{\HSSym{\{\mskip1.5mu} }\HSVar{o}\;\HSVar{r}\;\mathbin{\HSCon{:}}\;\HSCon{Record}\HSSpecial{\HSSym{\mskip1.5mu\}}}\;\mathrel{\HSSym{\to}} \;\HT{RecordChainFrom}\;\HSVar{o}\;\HSVar{r}\;\mathrel{\HSSym{\to}} \;\HSCon{Set}\;\HSKeyword{where}{}\<[E]%
\\
\>[B]{}\hsindent{4}{}\<[4]%
\>[4]{}\HV{\hbox{\it 0\guydash{}chain}}\;{}\<[15]%
\>[15]{}\mathbin{\HSCon{:}}\;\HSSym{\forall}\;\HSSpecial{\HSSym{\{\mskip1.5mu} }\HSVar{o}\;\HSVar{r}\HSSpecial{\HSSym{\mskip1.5mu\}}}\;\HSSpecial{\HSSym{\{\mskip1.5mu} }\HSVar{rc}\;\mathbin{\HSCon{:}}\;\HT{RecordChainFrom}\;\HSVar{o}\;\HSVar{r}\HSSpecial{\HSSym{\mskip1.5mu\}}}\;\mathrel{\HSSym{\to}} \;\HT{\mathbb{K}\hbox{\it \guydash{}chain}}\;\HSCon{R}\;\HSVar{0}\;\HSVar{rc}{}\<[E]%
\\
\>[B]{}\hsindent{4}{}\<[4]%
\>[4]{}\HV{\hbox{\it s\guydash{}chain}}\;{}\<[15]%
\>[15]{}\mathbin{\HSCon{:}}\;\HSSym{\forall}\;\HSSpecial{\HSSym{\{\mskip1.5mu} }\HSVar{k}\;\HSVar{o}\;\HSVar{r}\HSSpecial{\HSSym{\mskip1.5mu\}}}\;\HSSpecial{\HSSym{\{\mskip1.5mu} }\HSVar{rc}\;\mathbin{\HSCon{:}}\;\HT{RecordChainFrom}\;\HSVar{o}\;\HSVar{r}\HSSpecial{\HSSym{\mskip1.5mu\}}}\;\HSSpecial{\HSSym{\{\mskip1.5mu} }\HSVar{b}\;\mathbin{\HSCon{:}}\;\HSCon{Block}\HSSpecial{\HSSym{\mskip1.5mu\}}}\;\HSSpecial{\HSSym{\{\mskip1.5mu} }\HSVar{q}\;\mathbin{\HSCon{:}}\;\HSCon{QC}\HSSpecial{\HSSym{\mskip1.5mu\}}}{}\<[E]%
\\
\>[15]{}\mathrel{\HSSym{\to}} \;\HSSpecial{(}\HV{r\!\!\leftarrow\!\!b}\;\mathbin{\HSCon{:}}\;\HSVar{r}\;{}\<[31]%
\>[31]{}\HT{\leftarrow}\;\HSCon{B}\;\HSVar{b}\HSSpecial{)}\;\mathrel{\HSSym{\to}} \;\HSSpecial{(}\HSVar{prf}\;\mathbin{\HSCon{:}}\;\HSCon{R}\;\HSVar{k}\;\HSVar{r}\;\HSSpecial{(}\HSCon{B}\;\HSVar{b}\HSSpecial{)}\HSSpecial{)}{}\<[E]%
\\
\>[15]{}\mathrel{\HSSym{\to}} \;\HSSpecial{(}\HV{b\!\!\leftarrow\!\!q}\;\mathbin{\HSCon{:}}\;\HSCon{B}\;\HSVar{b}\;\HT{\leftarrow}\;\HSCon{Q}\;\HSVar{q}\HSSpecial{)}\;\mathrel{\HSSym{\to}} \;\HT{\mathbb{K}\hbox{\it \guydash{}chain}}\;\HSCon{R}\;\HSVar{k}\;\HSVar{rc}{}\<[E]%
\\
\>[15]{}\mathrel{\HSSym{\to}} \;\HT{\mathbb{K}\hbox{\it \guydash{}chain}}\;\HSCon{R}\;\HSSpecial{(}\HSVar{suc}\;\HSVar{k}\HSSpecial{)}\;\HSSpecial{(}\HSVar{step}\;\HSSpecial{(}\HSVar{step}\;\HSVar{rc}\;\HV{r\!\!\leftarrow\!\!b}\HSSpecial{)}\;\HV{b\!\!\leftarrow\!\!q}\HSSpecial{)}{}\<[E]%
\ColumnHook
\end{hscode}\resethooks
\end{myhs}

Block \ensuremath{\HSVar{\mathit{b}_{0}}} (and those preceding it) are committed
if \ensuremath{\HSVar{\mathit{b}_{0}}} is the head of a contiguous 3-chain: there is a \ensuremath{\HT{RecordChain}} that contains \ensuremath{\HSVar{b}} followed by
blocks \ensuremath{\HSVar{\mathit{b}_{1}}} and \ensuremath{\HSVar{\mathit{b}_{2}}}, such that the rounds of blocks \ensuremath{\HSVar{\mathit{b}_{0}}}, \ensuremath{\HSVar{\mathit{b}_{1}}} and \ensuremath{\HSVar{\mathit{b}_{2}}} are consecutive.
This is called
a \ensuremath{\HSCon{CommitRule}} %
(\ensuremath{\HSVar{kchainBlock}\;\HSVar{n}\;\HSVar{\mathit{c}_{3}}} is the \ensuremath{\HSVar{n}}th \ensuremath{\HSCon{Block}} from the end of \ensuremath{\HSVar{\mathit{c}_{3}}}):

\begin{myhs}
\begin{hscode}\SaveRestoreHook
\column{B}{@{}>{\hspre}l<{\hspost}@{}}%
\column{4}{@{}>{\hspre}l<{\hspost}@{}}%
\column{16}{@{}>{\hspre}l<{\hspost}@{}}%
\column{23}{@{}>{\hspre}l<{\hspost}@{}}%
\column{E}{@{}>{\hspre}l<{\hspost}@{}}%
\>[B]{}\HSKeyword{data}\;\HSCon{CommitRuleFrom}\;{}\<[23]%
\>[23]{}\HSSpecial{\HSSym{\{\mskip1.5mu} }\HSVar{o}\;\HSVar{r}\;\mathbin{\HSCon{:}}\;\HSCon{Record}\HSSpecial{\HSSym{\mskip1.5mu\}}}\;{}\<[E]%
\\
\>[23]{}\HSSpecial{(}\HSVar{rc}\;\mathbin{\HSCon{:}}\;\HT{RecordChainFrom}\;\HSVar{o}\;\HSVar{r}\HSSpecial{)}\;\HSSpecial{(}\HSVar{b}\;\mathbin{\HSCon{:}}\;\HSCon{Block}\HSSpecial{)}\;\mathbin{\HSCon{:}}\;\HSCon{Set}\;\HSKeyword{where}{}\<[E]%
\\
\>[B]{}\hsindent{4}{}\<[4]%
\>[4]{}\HV{\hbox{\it commit\guydash{}rule}}\;{}\<[16]%
\>[16]{}\mathbin{\HSCon{:}}\;\HSSpecial{(}\HSVar{\mathit{c}_{3}}\;\mathbin{\HSCon{:}}\;\HT{\mathbb{K}\hbox{\it \guydash{}chain}}\;\HSCon{Contig}\;\HSVar{3}\;\HSVar{rc}\HSSpecial{)}\;\mathrel{\HSSym{\to}} \;\HSVar{b}\;\mathrel{\HSSym{\equiv}} \;\HSVar{kchainBlock}\;\HSVar{2}\;\HSVar{\mathit{c}_{3}}\;{}\<[E]%
\\
\>[16]{}\mathrel{\HSSym{\to}} \;\HSCon{CommitRuleFrom}\;\HSVar{rc}\;\HSVar{b}{}\<[E]%
\ColumnHook
\end{hscode}\resethooks
\end{myhs}

\subsection{First correctness property: \ensuremath{\HSVar{thmS5}}}
\label{sec:thmsfive}
We can now explain the first high-level property we prove for our abstract model,
\ensuremath{\HSVar{thmS5}}.  (Because our work has been influenced by versions of the \hotstuff{}~\cite{hotstuff-podc-19}
and \librabft{} papers~\cite{libra-2019-06-28,libra-2020-05-26}, some of our properties are named
after properties presented informally in those papers.  For example, \ensuremath{\HSVar{thmS5}} is named
after Theorem S5 in~\cite{libra-2019-06-28}.)
It states that, if two blocks \ensuremath{\HSVar{b}} and \ensuremath{\HSVar{b'}} are committed via
\ensuremath{\HSCon{CommitRule}\;\HSVar{rc}\;\HSVar{b}} and \ensuremath{\HSCon{CommitRule}\;\HV{rc^{\prime}}\;\HV{b^{\prime}}},
respectively, then one of the blocks is contained in the record chain of the other.
This property ensures that all
committed \ensuremath{\HSCon{Block}}s are on a single non-branching path in the tree of \ensuremath{\HSCon{Record}}s.

\begin{myhs}
\begin{hscode}\SaveRestoreHook
\column{B}{@{}>{\hspre}l<{\hspost}@{}}%
\column{11}{@{}>{\hspre}l<{\hspost}@{}}%
\column{22}{@{}>{\hspre}l<{\hspost}@{}}%
\column{39}{@{}>{\hspre}l<{\hspost}@{}}%
\column{50}{@{}>{\hspre}l<{\hspost}@{}}%
\column{E}{@{}>{\hspre}l<{\hspost}@{}}%
\>[B]{}\HSVar{thmS5}\;{}\<[11]%
\>[11]{}\mathbin{\HSCon{:}}\;\HSSym{\forall}\;\HSSpecial{\HSSym{\{\mskip1.5mu} }\HSVar{q}\;\HV{q^{\prime}}\HSSpecial{\HSSym{\mskip1.5mu\}}}\;\mathrel{\HSSym{\to}} \;\HSSpecial{\HSSym{\{\mskip1.5mu} }\HSVar{rc}\;{}\<[39]%
\>[39]{}\mathbin{\HSCon{:}}\;\HT{RecordChain}\;\HSSpecial{(}\HSCon{Q}\;\HSVar{q}\HSSpecial{)}\HSSpecial{\HSSym{\mskip1.5mu\}}}\;\mathrel{\HSSym{\to}} \;\HSCon{AllInSys}\;\HSVar{rc}{}\<[E]%
\\
\>[11]{}\mathrel{\HSSym{\to}} \;\HSSpecial{\HSSym{\{\mskip1.5mu} }\HV{rc^{\prime}}\;\mathbin{\HSCon{:}}\;\HT{RecordChain}\;\HSSpecial{(}\HSCon{Q}\;\HV{q^{\prime}}\HSSpecial{)}\HSSpecial{\HSSym{\mskip1.5mu\}}}\;{}\<[50]%
\>[50]{}\mathrel{\HSSym{\to}} \;\HSCon{AllInSys}\;\HV{rc^{\prime}}{}\<[E]%
\\
\>[11]{}\mathrel{\HSSym{\to}} \;\HSSpecial{\HSSym{\{\mskip1.5mu} }\HSVar{b}\;\HV{b^{\prime}}\;\mathbin{\HSCon{:}}\;\HSCon{Block}\HSSpecial{\HSSym{\mskip1.5mu\}}}\;\mathrel{\HSSym{\to}} \;\HSCon{CommitRule}\;\HSVar{rc}\;\HSVar{b}\;\mathrel{\HSSym{\to}} \;\HSCon{CommitRule}\;\HV{rc^{\prime}}\;\HV{b^{\prime}}{}\<[E]%
\\
\>[11]{}\mathrel{\HSSym{\to}} \;\HSCon{Either}\;{}\<[22]%
\>[22]{}\HT{\hbox{\it NonInjective\guydash{}$\equiv$}}\;\HSSpecial{(}\HSCon{Either}\;\HSSpecial{(}\HSSpecial{(}\HSCon{B}\;\HSVar{b}\HSSpecial{)}\;\mathbin{\HT{\in\!\!RC}}\;\HV{rc^{\prime}}\HSSpecial{)}\;\HSSpecial{(}\HSSpecial{(}\HSCon{B}\;\HV{b^{\prime}}\HSSpecial{)}\;\mathbin{\HT{\in\!\!RC}}\;\HSVar{rc}\HSSpecial{)}\HSSpecial{)}{}\<[E]%
\ColumnHook
\end{hscode}\resethooks
\end{myhs}

\ensuremath{\HSCon{AllInSys}\;\HSVar{rc}} means that each record in \ensuremath{\HSVar{rc}} is ``in'' the abstract system
according to an implementation-specific predicate over abstract \ensuremath{\HSCon{Record}}s called \ensuremath{\HSCon{InSys}}, which is provided
as a module parameter.  For  purposes of \ensuremath{\HSCon{AllInSys}}, a record \ensuremath{\HSVar{r}} being ``in'' a record chain \ensuremath{\HSVar{rc}}
is captured by a simple recursive definition: if \ensuremath{\HSVar{rc}} is formed by extending
record chain \ensuremath{\HV{rc^{\prime}}} by record \ensuremath{\HV{r^{\prime}}}, then \ensuremath{\HSVar{r}} is ``in'' \ensuremath{\HSVar{rc}} iff
\ensuremath{\HSVar{r}\;\mathrel{\HSSym{=}}\;\HV{r^{\prime}}} or \ensuremath{\HSVar{r}} is ``in'' \ensuremath{\HV{rc^{\prime}}}.
On the other hand, as explained in~\Cref{sec:abstract:proof},
\ensuremath{\mathbin{\HT{\in\!\!RC}}} represents a more complicated notion of a record being ``in'' a record chain.

Note that \ensuremath{\HSVar{thmS5}} requires that \emph{either} \ensuremath{\HT{\hbox{\it NonInjective\guydash{}$\equiv$}}} holds \emph{or} one of the committed \ensuremath{\HSCon{Block}}s is in a \ensuremath{\HT{RecordChain}} ending at the other.
The \ensuremath{\HT{\hbox{\it NonInjective\guydash{}$\equiv$}}} disjunct---which is shared by many of the properties discussed below---reflects
that we prove \ensuremath{\HSVar{thmS5}} modulo injectivity of \ensuremath{\HSCon{Block}}
ids, as discussed above.

In~\Cref{sec:hashinj},
we explain how we refine the definition of \ensuremath{\HSVar{thmS5}} and other properties
in order to relate our abstract proofs to the security properties of a
concrete implementation that is proved correct using
them.  For now, however, we can think of the following simplified definition of \ensuremath{\HT{\hbox{\it NonInjective\guydash{}$\equiv$}}}:

\begin{myhs}
\begin{hscode}\SaveRestoreHook
\column{B}{@{}>{\hspre}l<{\hspost}@{}}%
\column{25}{@{}>{\hspre}l<{\hspost}@{}}%
\column{E}{@{}>{\hspre}l<{\hspost}@{}}%
\>[B]{}\HT{\hbox{\it NonInjective\guydash{}$\equiv$}}\;\mathbin{\HSCon{:}}\;\HSCon{Set}{}\<[E]%
\\
\>[B]{}\HT{\hbox{\it NonInjective\guydash{}$\equiv$}}\;\mathrel{\HSSym{=}}\;\HTNI{\Sigma}\;{}\<[25]%
\>[25]{}\HSSpecial{(}\HSCon{Block}\;\mathbin{\HT{\times}}\;\HSCon{Block}\HSSpecial{)}\;{}\<[E]%
\\
\>[25]{}\HSSpecial{(}\HSSym{\lambda} \;\HSSpecial{\HSSym{\{\mskip1.5mu} }\HSSpecial{(}\HSVar{\mathit{b}_{0}}\;\mathbin{\HSSym{,}}\;\HSVar{\mathit{b}_{1}}\HSSpecial{)}\;\mathrel{\HSSym{\to}} \;\HSVar{\mathit{b}_{0}}\;\mathrel{\HSSym{\not\equiv}} \;\HSVar{\mathit{b}_{1}}\;\mathbin{\HT{\times}}\;\HSVar{bId}\;\HSVar{\mathit{b}_{0}}\;\mathrel{\HSSym{\equiv}} \;\HSVar{bId}\;\HSVar{\mathit{b}_{1}}\HSSpecial{\HSSym{\mskip1.5mu\}}}\HSSpecial{)}{}\<[E]%
\ColumnHook
\end{hscode}\resethooks
\end{myhs}

The \ensuremath{\HTNI{\Sigma}} notation is similar to the \ensuremath{\HT{\exists[}{\cdot }\HT{]}{\cdot }} notation introduced earlier, except that it specifies the \emph{type} of the existentially
quantified value (not just a name, as with \ensuremath{\HT{\exists[}{\cdot }\HT{]}{\cdot }}) and the condition on the value of that
type is expressed as a predicate on that type.  Thus, a value of type \ensuremath{\HT{\hbox{\it NonInjective\guydash{}$\equiv$}}} comprises
a \emph{pair} of (abstract)
\ensuremath{\HSCon{Block}}s---\ensuremath{\HSVar{\mathit{b}_{0}}} and \ensuremath{\HSVar{\mathit{b}_{1}}}---that
are \emph{different} but have the \emph{same} id.

\subsection{Precisely defining protocol rules}
\label{sec:abstract:rules}

Module \ensuremath{\HSCon{LibraBFT.Abstract.RecordChain.Properties}}
contains the proof of \ensuremath{\HSVar{thmS5}},
which requires module parameters representing assumptions about \ensuremath{\HSCon{Record}}s that are \ensuremath{\HSCon{InSys}}.
These
assumptions capture the key properties that an implementation must ensure.
Part of our contribution is to precisely define these assumptions
in an abstract way, independent of any particular implementation.

Implementations described in various papers~\cite{hotstuff-podc-19,libra-2019-06-28,libra-2020-05-26}
are all based on the same core ideas, but
 differ substantially in detail.  None of these papers gives a precise definition of the core protocol.
 Early versions of the \librabft{} papers~\cite{libra-2019-06-28} come closest, providing explicit statements of two
 ``voting constraints''.

These voting constraints (``Increasing Round'' and ``Preferred Round'') were a starting point for us,
but they are not entirely suitable for our purposes.
For example, the ``Increasing Round''
constraint is originally stated as: \emph{
An honest node that voted once for \ensuremath{\HSCon{B}} in the past may only vote for \ensuremath{\HSCon{B'}} if \ensuremath{\HSVar{round}\;\HSSpecial{(}\HSCon{B}\HSSpecial{)}\;\HSVar{<}\;\HSVar{round}\;\HSSpecial{(}\HSCon{B'}\HSSpecial{)}}.}
However, to interpret this as a protocol rule, we would need to define precisely what it means to have ``voted in the past''\!.
Our proof efforts revealed that it suffices to require
that an honest peer does not sign and send \emph{different} (abstract) votes for the same round
(regardless of order)\extendedalt{. This can be stated as:}{:}

\begin{minipage}[t]{5.5cm}
\begin{myhs}
\begin{hscode}\SaveRestoreHook
\column{B}{@{}>{\hspre}l<{\hspost}@{}}%
\column{20}{@{}>{\hspre}l<{\hspost}@{}}%
\column{27}{@{}>{\hspre}l<{\hspost}@{}}%
\column{E}{@{}>{\hspre}l<{\hspost}@{}}%
\>[B]{}\HSCon{VotesOnlyOnceRule}\;\mathbin{\HSCon{:}}\;\HSCon{Set}\;\HVNI{\ell}{}\<[E]%
\\
\>[B]{}\HSCon{VotesOnlyOnceRule}\;{}\<[20]%
\>[20]{}\mathrel{\HSSym{=}}\;\HSSpecial{(}\HSVar{a}\;\mathbin{\HSCon{:}}\;\HSCon{Member}\HSSpecial{)}\;\mathrel{\HSSym{\to}} \;\HSCon{MetaHonestMember}\;\HSVar{a}\;{}\<[E]%
\\
\>[20]{}\mathrel{\HSSym{\to}} \;\HSSym{\forall}\;\HSSpecial{\HSSym{\{\mskip1.5mu} }\HSVar{q}\;\HSVar{q'}\HSSpecial{\HSSym{\mskip1.5mu\}}}\;\mathrel{\HSSym{\to}} \;\HSCon{InSys}\;\HSSpecial{(}\HSCon{Q}\;\HSVar{q}\HSSpecial{)}\;\mathrel{\HSSym{\to}} \;\HSCon{InSys}\;\HSSpecial{(}\HSCon{Q}\;\HSVar{q'}\HSSpecial{)}{}\<[E]%
\\
\>[20]{}\mathrel{\HSSym{\to}} \;\HSSpecial{(}\HSVar{v}\;{}\<[27]%
\>[27]{}\mathbin{\HSCon{:}}\;\HSVar{a}\;\mathbin{\HT{\in\!\!QC}}\;\HSVar{q}\HSSpecial{)}\;\HSSpecial{(}\HSVar{v'}\;\mathbin{\HSCon{:}}\;\HSVar{a}\;\mathbin{\HT{\in\!\!QC}}\;\HSVar{q'}\HSSpecial{)}{}\<[E]%
\\
\>[20]{}\mathrel{\HSSym{\to}} \;\HSVar{vRound}\;\HSSpecial{(}\mathbin{\HT{\in\!\!QC\hbox{\it \guydash{}Vote}}}\;\HSVar{q}\;\HSVar{v}\HSSpecial{)}\;\mathrel{\HSSym{\equiv}} \;\HSVar{vRound}\;\HSSpecial{(}\mathbin{\HT{\in\!\!QC\hbox{\it \guydash{}Vote}}}\;\HSVar{q'}\;\HSVar{v'}\HSSpecial{)}{}\<[E]%
\\
\>[20]{}\mathrel{\HSSym{\to}} \;\mathbin{\HT{\in\!\!QC\hbox{\it \guydash{}Vote}}}\;\HSVar{q}\;\HSVar{v}\;\mathrel{\HSSym{\equiv}} \;\mathbin{\HT{\in\!\!QC\hbox{\it \guydash{}Vote}}}\;\HSVar{q'}\;\HSVar{v'}{}\<[E]%
\ColumnHook
\end{hscode}\resethooks
\end{myhs}
\end{minipage}

For generality,
\ensuremath{\HSCon{InSys}} is assumed to return a type from some arbitrary universe \cite{wikipedia-universes} with level \ensuremath{\HVNI{\ell}}.
The \ensuremath{\HSVar{v}} parameter is evidence that there is a \ensuremath{\HSCon{Vote}} by member \ensuremath{\HSVar{a}} represented in \ensuremath{\HSVar{q}} (a \ensuremath{\HSCon{QC}}),
and \ensuremath{\mathbin{\HT{\in\!\!QC\hbox{\it \guydash{}Vote}}}\;\HSVar{q}\;\HSVar{v}} is that (abstract) \ensuremath{\HSCon{Vote}}.  Thus,
\ensuremath{\HSCon{VotesOnlyOnceRule}} requires that, if there are two \ensuremath{\HSCon{Vote}}s for the same round by an honest member \ensuremath{\HSVar{a}} in \ensuremath{\HSCon{QC}}s in the system,
then the \ensuremath{\HSCon{Vote}}s are equal.

The second constraint---\ensuremath{\HSCon{PreferredRoundRule}}---is more complicated.  It is \linebreak based on
the voting constraint called ``Locked Round'' in early versions of the LibraBFT paper~\cite{libra-2019-06-28};
similar constraints on voting are followed by \hotstuff{}~\cite{hotstuff-podc-19}
and by later versions of \librabft~\cite{libra-2020-05-26}.
The essence of this rule is that, if an honest peer contributes a
\ensuremath{\HSCon{Vote}} to \ensuremath{\HSVar{q}} (a \ensuremath{\HSCon{QC}}) that commits a \ensuremath{\HSCon{Block}} (\ensuremath{\HSVar{\mathit{c}_{3}}} is essentially a \ensuremath{\HSCon{CommitRule}} that commits the
\ensuremath{\HSCon{Block}} identified by \ensuremath{\HSVar{kchainBlock}\;\HSVar{2}\;\HSVar{\mathit{c}_{3}}}), then it does not vote in a higher
round for a \ensuremath{\HSCon{Block}} unless the round of the \emph{previous} \ensuremath{\HSCon{Block}} is at least that of the committed \ensuremath{\HSCon{Block}}.
This is a key requirement to avoid voting to commit another \ensuremath{\HSCon{Block}} that conflicts with the first.

\begin{myhs}
\begin{hscode}\SaveRestoreHook
\column{B}{@{}>{\hspre}l<{\hspost}@{}}%
\column{3}{@{}>{\hspre}l<{\hspost}@{}}%
\column{14}{@{}>{\hspre}l<{\hspost}@{}}%
\column{25}{@{}>{\hspre}l<{\hspost}@{}}%
\column{E}{@{}>{\hspre}l<{\hspost}@{}}%
\>[B]{}\HSCon{PreferredRoundRule}\;\mathbin{\HSCon{:}}\;\HSCon{Set}\;\HVNI{\ell}{}\<[E]%
\\
\>[B]{}\HSCon{PreferredRoundRule}\;{}\<[E]%
\\
\>[B]{}\hsindent{3}{}\<[3]%
\>[3]{}\mathrel{\HSSym{=}}\;\HSSym{\forall}\;\HSVar{a}\;\HSSpecial{\HSSym{\{\mskip1.5mu} }\HSVar{q}\;\HSVar{q'}\HSSpecial{\HSSym{\mskip1.5mu\}}}\;\mathrel{\HSSym{\to}} \;\HSCon{MetaHonestMember}\;\HSVar{a}\;\mathrel{\HSSym{\to}} \;\HSCon{InSys}\;\HSSpecial{(}\HSCon{Q}\;\HSVar{q}\HSSpecial{)}\;\mathrel{\HSSym{\to}} \;\HSCon{InSys}\;\HSSpecial{(}\HSCon{Q}\;\HSVar{q'}\HSSpecial{)}{}\<[E]%
\\
\>[B]{}\hsindent{3}{}\<[3]%
\>[3]{}\mathrel{\HSSym{\to}} \;\HSSpecial{\HSSym{\{\mskip1.5mu} }\HSVar{rc}\;\mathbin{\HSCon{:}}\;\HT{RecordChain}\;\HSSpecial{(}\HSCon{Q}\;\HSVar{q}\HSSpecial{)}\HSSpecial{\HSSym{\mskip1.5mu\}}}\;\HSSpecial{\HSSym{\{\mskip1.5mu} }\HSVar{n}\;\mathbin{\HSCon{:}}\;\HT{\mathbb{N}}\HSSpecial{\HSSym{\mskip1.5mu\}}}\;\mathrel{\HSSym{\to}} \;\HSSpecial{(}\HSVar{\mathit{c}_{3}}\;\mathbin{\HSCon{:}}\;\HT{\mathbb{K}\hbox{\it \guydash{}chain}}\;\HSCon{Contig}\;\HSSpecial{(}\HSVar{3}\;\mathbin{\HSSym{+}}\;\HSVar{n}\HSSpecial{)}\;\HSVar{rc}\HSSpecial{)}{}\<[E]%
\\
\>[B]{}\hsindent{3}{}\<[3]%
\>[3]{}\mathrel{\HSSym{\to}} \;\HSSpecial{(}\HSVar{v}\;\mathbin{\HSCon{:}}\;\HSVar{a}\;\mathbin{\HT{\in\!\!QC}}\;\HSVar{q}\HSSpecial{)}\;\HSSpecial{(}\HSVar{rc'}\;\mathbin{\HSCon{:}}\;\HT{RecordChain}\;\HSSpecial{(}\HSCon{Q}\;\HSVar{q'}\HSSpecial{)}\HSSpecial{)}\;\HSSpecial{(}\HSVar{v'}\;\mathbin{\HSCon{:}}\;\HSVar{a}\;\mathbin{\HT{\in\!\!QC}}\;\HSVar{q'}\HSSpecial{)}{}\<[E]%
\\
\>[B]{}\hsindent{3}{}\<[3]%
\>[3]{}\mathrel{\HSSym{\to}} \;\HSVar{vRound}\;\HSSpecial{(}\mathbin{\HT{\in\!\!QC\hbox{\it \guydash{}Vote}}}\;\HSVar{q}\;\HSVar{v}\HSSpecial{)}\;\HSVar{<}\;\HSVar{vRound}\;\HSSpecial{(}\mathbin{\HT{\in\!\!QC\hbox{\it \guydash{}Vote}}}\;\HSVar{q'}\;\HSVar{v'}\HSSpecial{)}{}\<[E]%
\\
\>[B]{}\hsindent{3}{}\<[3]%
\>[3]{}\mathrel{\HSSym{\to}} \;\HSCon{Either}\;{}\<[14]%
\>[14]{}\HT{\hbox{\it NonInjective\guydash{}$\equiv$}}\;{}\<[E]%
\\
\>[14]{}\HSSpecial{(}\HSVar{getRound}\;{}\<[25]%
\>[25]{}\HSSpecial{(}\HSVar{kchainBlock}\;\HSSpecial{(}\HSVar{suc}\;\HSSpecial{(}\HSVar{suc}\;\HSVar{zero}\HSSpecial{)}\HSSpecial{)}\;\HSVar{\mathit{c}_{3}}\HSSpecial{)}\;\mathrel{\HSSym{\leq}} \;\HSVar{prevRound}\;\HSVar{rc'}\HSSpecial{)}{}\<[E]%
\ColumnHook
\end{hscode}\resethooks
\end{myhs}

\extended{
Interestingly, \librabft{} updates the \texttt{preferred\_round}
variable used to comply with this rule even if a \ensuremath{\HT{\hbox{\it 3\guydash{}chain}}} resulting from
including the vote in a \ensuremath{\HSCon{QC}} would not have
contiguous rounds (and thus would not result in a \ensuremath{\HSCon{CommitRule}}).  Thus, our development shows
that the implementation could be less conservative without violating correctness.
}

\subsection{The proof of \ensuremath{\HSVar{thmS5}}}
\label{sec:abstract:proof}

Our proof of \ensuremath{\HSVar{thmS5}} is similar to
the manual proof presented an early version of the
\librabft{} paper~\cite{libra-2019-06-28}.
However, a formal, machine-checked proof must address
many details that are glossed over in the manual proof.  Furthermore,
as discussed in~\Cref{sec:abstract:rules}, making our
assumptions about honest peers' \ensuremath{\HSCon{Vote}}s precise and
implementation-independent required somewhat
different assumptions.

To help the reader approach the formal, machine-checked proofs in our
open-source development~\cite{librabft-agda},
we describe below some of its key proofs and properties.

We first introduce two key lemmas.
Roughly speaking, \ensuremath{\HV{lemmaS2}} states that there can be at most one certified \ensuremath{\HSCon{Block}} per round.
Its proof
depends on the \ensuremath{\HV{\hbox{\it bft\guydash{}assumption}}}: for two \ensuremath{\HSCon{QC}}s, there is some honest peer with
\ensuremath{\HSCon{Vote}}s in each.  By the assumption that honest peers obey \ensuremath{\HSCon{VotesOnlyOnceRule}},
if the blocks certified by the two \ensuremath{\HSCon{QC}}s have the same round, then both \ensuremath{\HSCon{Vote}}s are for the same \ensuremath{\HSCon{BlockId}}.
However, this does \emph{not} imply the \ensuremath{\HSCon{QC}}s certify the same \ensuremath{\HSCon{Block}}.
For this reason, the conclusion of \ensuremath{\HSVar{lemmaS2}} is that \emph{either} \ensuremath{\HSVar{bId}} is non-injective \emph{or} \ensuremath{\HSVar{\mathit{b}_{0}}\;\mathrel{\HSSym{\equiv}} \;\HSVar{\mathit{b}_{1}}}.

\begin{myhs}
\begin{hscode}\SaveRestoreHook
\column{B}{@{}>{\hspre}l<{\hspost}@{}}%
\column{10}{@{}>{\hspre}l<{\hspost}@{}}%
\column{E}{@{}>{\hspre}l<{\hspost}@{}}%
\>[B]{}\HSVar{lemmaS2}\;{}\<[10]%
\>[10]{}\mathbin{\HSCon{:}}\;\HSSym{\forall}\;\HSSpecial{\HSSym{\{\mskip1.5mu} }\HSVar{\mathit{b}_{0}}\;\HSVar{\mathit{b}_{1}}\;\mathbin{\HSCon{:}}\;\HSCon{Block}\HSSpecial{\HSSym{\mskip1.5mu\}}}\;\HSSpecial{\HSSym{\{\mskip1.5mu} }\HSVar{\mathit{q}_{0}}\;\HSVar{\mathit{q}_{1}}\;\mathbin{\HSCon{:}}\;\HSCon{QC}\HSSpecial{\HSSym{\mskip1.5mu\}}}\;\mathrel{\HSSym{\to}} \;\HSCon{InSys}\;\HSSpecial{(}\HSCon{Q}\;\HSVar{\mathit{q}_{0}}\HSSpecial{)}\;\mathrel{\HSSym{\to}} \;\HSCon{InSys}\;\HSSpecial{(}\HSCon{Q}\;\HSVar{\mathit{q}_{1}}\HSSpecial{)}{}\<[E]%
\\
\>[10]{}\mathrel{\HSSym{\to}} \;\HSSpecial{(}\HSVar{\mathit{p}_{0}}\;\mathbin{\HSCon{:}}\;\HSCon{B}\;\HSVar{\mathit{b}_{0}}\;\HT{\leftarrow}\;\HSCon{Q}\;\HSVar{\mathit{q}_{0}}\HSSpecial{)}\;\HSSpecial{(}\HSVar{\mathit{p}_{1}}\;\mathbin{\HSCon{:}}\;\HSCon{B}\;\HSVar{\mathit{b}_{1}}\;\HT{\leftarrow}\;\HSCon{Q}\;\HSVar{\mathit{q}_{1}}\HSSpecial{)}{}\<[E]%
\\
\>[10]{}\mathrel{\HSSym{\to}} \;\HSVar{getRound}\;\HSVar{\mathit{b}_{0}}\;\mathrel{\HSSym{\equiv}} \;\HSVar{getRound}\;\HSVar{\mathit{b}_{1}}{}\<[E]%
\\
\>[10]{}\mathrel{\HSSym{\to}} \;\HSCon{Either}\;\HT{\hbox{\it NonInjective\guydash{}$\equiv$}}\;\HSSpecial{(}\HSVar{\mathit{b}_{0}}\;\mathrel{\HSSym{\equiv}} \;\HSVar{\mathit{b}_{1}}\HSSpecial{)}{}\<[E]%
\ColumnHook
\end{hscode}\resethooks
\end{myhs}

Similarly, \ensuremath{\HSVar{lemmaS3}} makes the \ensuremath{\HSCon{PreferredRoundRule}} apply to \ensuremath{\HSCon{QC}}s.

\begin{myhs}
\begin{hscode}\SaveRestoreHook
\column{B}{@{}>{\hspre}l<{\hspost}@{}}%
\column{10}{@{}>{\hspre}l<{\hspost}@{}}%
\column{21}{@{}>{\hspre}l<{\hspost}@{}}%
\column{47}{@{}>{\hspre}l<{\hspost}@{}}%
\column{E}{@{}>{\hspre}l<{\hspost}@{}}%
\>[B]{}\HSVar{lemmaS3}\;{}\<[10]%
\>[10]{}\mathbin{\HSCon{:}}\;\HSSym{\forall}\;\HSSpecial{\HSSym{\{\mskip1.5mu} }\HSVar{\mathit{r}_{2}}\;\HV{q^{\prime}}\HSSpecial{\HSSym{\mskip1.5mu\}}}\;\HSSpecial{\HSSym{\{\mskip1.5mu} }\HSVar{rc}\;\mathbin{\HSCon{:}}\;\HT{RecordChain}\;\HSVar{\mathit{r}_{2}}\HSSpecial{\HSSym{\mskip1.5mu\}}}\;\mathrel{\HSSym{\to}} \;\HSCon{InSys}\;\HSVar{\mathit{r}_{2}}{}\<[E]%
\\
\>[10]{}\mathrel{\HSSym{\to}} \;\HSSpecial{(}\HV{rc^{\prime}}\;\mathbin{\HSCon{:}}\;\HT{RecordChain}\;\HSSpecial{(}\HSCon{Q}\;\HV{q^{\prime}}\HSSpecial{)}\HSSpecial{)}\;\mathrel{\HSSym{\to}} \;\HSCon{InSys}\;\HSSpecial{(}\HSCon{Q}\;\HV{q^{\prime}}\HSSpecial{)}{}\<[E]%
\\
\>[10]{}\mathrel{\HSSym{\to}} \;\HSSpecial{(}\HSVar{\mathit{c}_{3}}\;\mathbin{\HSCon{:}}\;\HSVar{kchain}\;\HSCon{Contig}\;\HSVar{3}\;\HSVar{rc}\HSSpecial{)}\;\mathrel{\HSSym{\to}} \;\HSVar{round}\;\HSVar{\mathit{r}_{2}}\;\HSVar{<}\;\HSVar{getRound}\;\HV{q^{\prime}}{}\<[E]%
\\
\>[10]{}\mathrel{\HSSym{\to}} \;\HSCon{Either}\;{}\<[21]%
\>[21]{}\HT{\hbox{\it NonInjective\guydash{}$\equiv$}}\;\HSSpecial{(}\HSVar{getRound}\;{}\<[47]%
\>[47]{}\HSSpecial{(}\HSVar{kchainBlock}\;\HSSpecial{(}\HSVar{suc}\;\HSSpecial{(}\HSVar{suc}\;\HSVar{zero}\HSSpecial{)}\HSSpecial{)}\;\HSVar{\mathit{c}_{3}}\HSSpecial{)}\;{}\<[E]%
\\
\>[47]{}\mathrel{\HSSym{\leq}} \;\HSVar{prevRound}\;\HV{rc^{\prime}}\HSSpecial{)}{}\<[E]%
\ColumnHook
\end{hscode}\resethooks
\end{myhs}

The proof of \ensuremath{\HSVar{thmS5}} depends on a non-symmetric variant of it called \ensuremath{\HSVar{propS4}}:

\begin{myhs}
\begin{hscode}\SaveRestoreHook
\column{B}{@{}>{\hspre}l<{\hspost}@{}}%
\column{9}{@{}>{\hspre}l<{\hspost}@{}}%
\column{20}{@{}>{\hspre}l<{\hspost}@{}}%
\column{22}{@{}>{\hspre}l<{\hspost}@{}}%
\column{E}{@{}>{\hspre}l<{\hspost}@{}}%
\>[B]{}\HSVar{propS4}\;{}\<[9]%
\>[9]{}\mathbin{\HSCon{:}}\;\HSSym{\forall}\;\HSSpecial{\HSSym{\{\mskip1.5mu} }\HSVar{q}\;\HV{q^{\prime}}\HSSpecial{\HSSym{\mskip1.5mu\}}}\;\HSSpecial{\HSSym{\{\mskip1.5mu} }\HSVar{rc}\;\mathbin{\HSCon{:}}\;\HT{RecordChain}\;\HSSpecial{(}\HSCon{Q}\;\HSVar{q}\HSSpecial{)}\HSSpecial{\HSSym{\mskip1.5mu\}}}\;\mathrel{\HSSym{\to}} \;\HSCon{AllInSys}\;\HSVar{rc}{}\<[E]%
\\
\>[9]{}\mathrel{\HSSym{\to}} \;\HSSpecial{(}\HV{rc^{\prime}}\;\mathbin{\HSCon{:}}\;\HT{RecordChain}\;\HSSpecial{(}\HSCon{Q}\;\HV{q^{\prime}}\HSSpecial{)}\HSSpecial{)}\;\mathrel{\HSSym{\to}} \;\HSCon{AllInSys}\;\HV{rc^{\prime}}{}\<[E]%
\\
\>[9]{}\mathrel{\HSSym{\to}} \;\HSSpecial{(}\HSVar{\mathit{c}_{3}}\;\mathbin{\HSCon{:}}\;\HT{\mathbb{K}\hbox{\it \guydash{}chain}}\;\HSCon{Contig}\;\HSVar{3}\;\HSVar{rc}\HSSpecial{)}{}\<[E]%
\\
\>[9]{}\mathrel{\HSSym{\to}} \;\HSVar{getRound}\;{}\<[22]%
\>[22]{}\HSSpecial{(}\HSVar{kchainBlock}\;\HSSpecial{(}\HSVar{suc}\;\HSSpecial{(}\HSVar{suc}\;\HSVar{zero}\HSSpecial{)}\HSSpecial{)}\;\HSVar{\mathit{c}_{3}}\HSSpecial{)}\;\mathrel{\HSSym{\leq}} \;\HSVar{getRound}\;\HV{q^{\prime}}{}\<[E]%
\\
\>[9]{}\mathrel{\HSSym{\to}} \;\HSCon{Either}\;{}\<[20]%
\>[20]{}\HT{\hbox{\it NonInjective\guydash{}$\equiv$}}\;\HSSpecial{(}\HSCon{B}\;\HSSpecial{(}\HSVar{kchainBlock}\;\HSSpecial{(}\HSVar{suc}\;\HSSpecial{(}\HSVar{suc}\;\HSVar{zero}\HSSpecial{)}\HSSpecial{)}\;\HSVar{\mathit{c}_{3}}\HSSpecial{)}\;\mathbin{\HT{\in\!\!RC}}\;\HV{rc^{\prime}}\HSSpecial{)}{}\<[E]%
\ColumnHook
\end{hscode}\resethooks
\end{myhs}

Recall that \ensuremath{\mathbin{\HT{\in\!\!RC}}} is a specific representation of what it means for a \ensuremath{\HSCon{Record}} to be ``in'' a \ensuremath{\HT{RecordChain}}
that is precisely defined later, and note that \ensuremath{\HSVar{\mathit{c}_{3}}} is a \linebreak \ensuremath{\HT{\mathbb{K}\hbox{\it \guydash{}chain}}\;\HSCon{Contig}\;\HSVar{3}\;\HSVar{rc}}, for some \ensuremath{\HSVar{rc}}, i.e., a \ensuremath{\HSCon{CommitRule}}.

Proof overviews for \ensuremath{\HSVar{thmS5}} and \ensuremath{\HSVar{propS4}} are\extendedalt{ below.}{ in the extended paper~\cite{librabft-agda-extended}.}

\extended{
\paragraph{\bf Overview of proof for \ensuremath{\HSVar{thmS5}}:}
The proof of \ensuremath{\HSVar{thmS5}} can be found in \linebreak \ensuremath{\HSCon{LibraBFT.Abstract.RecordChain.Properties}}.
It constructs arguments for \ensuremath{\HSVar{propS4}} and establishes the
proof obligation using case analysis on its return value.
As with Lemmas \ensuremath{\HV{lemmaS2}} and \ensuremath{\HV{lemmaS3}}, \ensuremath{\HSVar{propS4}} returns either
evidence that the \ensuremath{\HSVar{bId}} function is not injective, or the desired property; in the former case, \ensuremath{\HSVar{thmS5}}
returns the evidence of the injectivity failure too.

The arguments for \ensuremath{\HSVar{propS4}} are determined by comparing the rounds of the \ensuremath{\HSCon{Block}}s
committed by the two \ensuremath{\HSCon{CommitRule}}s provided to \ensuremath{\HSVar{thmS5}}, and choosing the one with the smaller round (call it the first committed \ensuremath{\HSCon{Block}})
to provide to \ensuremath{\HSVar{propS4}} as the required \ensuremath{\HSCon{CommitRule}} \ensuremath{\HSVar{\mathit{c}_{3}}} (or the first if their rounds are equal).
The other record chain \ensuremath{\HV{rc^{\prime}}} provided to \ensuremath{\HSVar{propS4}} is derived from the other \ensuremath{\HSCon{CommitRule}}.  Evidence that the
round of the last record of this \ensuremath{\HT{RecordChain}} is at least that of the \ensuremath{\HSCon{Block}} committed by \ensuremath{\HSVar{\mathit{c}_{3}}} is determined by transitivity,
using the result of the above-mentioned comparison and a simple lemma \ensuremath{\HV{\hbox{\it kchain\guydash{}round\guydash{}$\leq$\guydash{}lemma$^{\prime}$}}};
this lemma inductively applies the constraints on the \ensuremath{\HT{\leftarrow}} components to show that the
round of the last \ensuremath{\HSCon{QC}} in the other \ensuremath{\HSCon{CommitRule}} is at least that of the \ensuremath{\HSCon{Block}} that it commits.\qed

\paragraph{\bf Overview of proof for \ensuremath{\HSVar{propS4}}:}
The proof of \ensuremath{\HSVar{propS4}} can also be found in \ensuremath{\HSCon{LibraBFT.Abstract.RecordChain.Properties}}.
It inductively removes a pair of records from the second record chain \ensuremath{\HV{rc^{\prime}}}, invoking itself recursively and
adding the pair of \ensuremath{\HSCon{Record}}s back to the result to construct final proof that the \ensuremath{\HSCon{Block}} that is committed by the provided \ensuremath{\HSCon{CommitRule}}
(that is, \ensuremath{\HT{\mathbb{K}\hbox{\it \guydash{}chain}}\;\HSCon{Config}\;\HSVar{3}\;\HSVar{rc}})---call it \ensuremath{\HSCon{B\char95 0}}---is ``in'' the second \ensuremath{\HT{RecordChain}}; we discuss \ensuremath{\mathbin{\HT{\in\!\!RC}}}---which represents a \ensuremath{\HSCon{Record}} being
``in'' a \ensuremath{\HT{RecordChain}}---below.  During this induction, \ensuremath{\HSVar{lemmaS3}} is used to determine that
the round of the last \ensuremath{\HSCon{Record}} \emph{after} the pair of \ensuremath{\HSCon{Record}}s is removed is at least that of \ensuremath{\HSCon{B\char95 0}}
(unless there is an injectivity failure), enabling the recursive call.
The result from \ensuremath{\HSVar{lemmaS3}} is also used in conjunction with
the fact that \ensuremath{\HSCon{Block}}s in a \ensuremath{\HT{RecordChain}} cannot have round 0 to ensure that we still have a \ensuremath{\HT{RecordChain}} ending
in a \ensuremath{\HSCon{QC}} after removing the pair of records, as is required for the recursive call to \ensuremath{\HSVar{propS4}}.

When the last \ensuremath{\HSCon{QC}} (\ensuremath{\HV{q^{\prime}}}) in \ensuremath{\HV{rc^{\prime}}} has a round that is at most
that of the last \ensuremath{\HSCon{Block}} in \ensuremath{\HSVar{\mathit{c}_{3}}}, we use \ensuremath{\HSVar{prop4base}},
which determines via case analysis which of the three \ensuremath{\HSCon{Block}}s in \ensuremath{\HSVar{\mathit{c}_{3}}} has the
same round as \ensuremath{\HV{q^{\prime}}}\!.  One of them---call it \ensuremath{\HSCon{B\char95 i}}---does due to the hypothesis that the round of
\ensuremath{\HV{q^{\prime}}} is at least that of the committed block \ensuremath{\HSCon{B\char95 0}}; \ensuremath{\HV{\hbox{\it propS4\guydash{}base\guydash{}lemma\guydash{}1}}} determines which \ensuremath{\HSCon{Block}} is \ensuremath{\HSCon{B\char95 i}}.
For each case, \ensuremath{\HV{\hbox{\it propS4\guydash{}base\guydash{}lemma\guydash{}2}}} is invoked with appropriate parameters, enabling the use of \ensuremath{\HSVar{lemmaS2}} to determine
that, because \ensuremath{\HSCon{B\char95 i}} has the same round as \ensuremath{\HV{q^{\prime}}} (and therefore of the \ensuremath{\HSCon{Block}} that precedes \ensuremath{\HV{q^{\prime}}} in \ensuremath{\HV{rc^{\prime}}}),
these are the same \ensuremath{\HSCon{Block}}.  This fact is used along with \ensuremath{\HT{\hbox{\it $\mathbb{K}$\guydash{}chain\guydash{}$\in\!$RC}}} to determine that \ensuremath{\HSCon{B\char95 0}} is ``in'' \ensuremath{\HV{rc^{\prime}}}, as required.\vspace*{.1in}\qed

}

Finally, we explain what it means for a \ensuremath{\HSCon{Block}} to be ``in'' a \ensuremath{\HT{RecordChain}},
as captured by the \ensuremath{\mathbin{\HT{\in\!\!RC}}} predicate.
It is tempting to think that, if \ensuremath{\HT{RecordChain}}s \ensuremath{\HSVar{rc}} and \ensuremath{\HV{rc^{\prime}}} both end at block \ensuremath{\HSVar{b}}, then
the requirements of \ensuremath{\HT{\_\!\!\ensuremath{\leftarrow}\!\!\_}} ensure that \ensuremath{\HSVar{rc}} and \ensuremath{\HV{rc^{\prime}}} are the
same \ensuremath{\HT{RecordChain}}.  However, suppose we have \ensuremath{\HSVar{q}\;\HT{\leftarrow}\;\HSVar{b}} and \ensuremath{\HV{q^{\prime}}\;\HT{\leftarrow}\;\HSVar{b}}, where \ensuremath{\HSVar{q}} and \ensuremath{\HV{q^{\prime}}} are \ensuremath{\HSCon{QC}}s.
The definition of \ensuremath{\HT{\_\!\!\ensuremath{\leftarrow}\!\!\_}} requires that \ensuremath{\HSVar{just}\;\HSSpecial{(}\HSVar{qCertBlockId}\;\HSVar{q}\HSSpecial{)}\;\mathrel{\HSSym{\equiv}} \;\HSVar{bprevQC}\;\HSVar{b}\;\mathrel{\HSSym{\equiv}} \;\HSVar{just}\;\HSSpecial{(}\HSVar{qCertBlockId}\;\HV{q^{\prime}}\HSSpecial{)}}.  This does \emph{not}
imply that \ensuremath{\HSVar{q}\;\mathrel{\HSSym{\equiv}} \;\HV{q^{\prime}}} because \ensuremath{\HSVar{q}} and \ensuremath{\HV{q^{\prime}}} may \extended{be two valid \ensuremath{\HSCon{QC}}s that} include different \ensuremath{\HSCon{Vote}}s,
reflecting the reality that two peers may be convinced to extend the same \ensuremath{\HSCon{Block}} by two \emph{different} valid \ensuremath{\HSCon{QC}}s.

Therefore, \extended{we cannot prove that \ensuremath{\HSVar{rc}\;\mathrel{\HSSym{\equiv}} \;\HV{rc^{\prime}}} just because both \ensuremath{\HT{RecordChain}}s end at the same \ensuremath{\HSCon{Record}}.  Instead,} we need a notion of \emph{equivalent}
\ensuremath{\HT{RecordChain}}s that contain the same \ensuremath{\HSCon{Block}}s and equivalent \ensuremath{\HSCon{QC}}s: two \ensuremath{\HSCon{QCs}} are equivalent iff they certify the same \ensuremath{\HSCon{Block}} (i.e,
their \ensuremath{\HSVar{qCertBlockId}} components are equal).  These notions are captured by \ensuremath{\mathrel{\HT{\approx\!\!\mathit{RC}}}} (defined in \ensuremath{\HSCon{LibraBFT.Abstract.RecordChain}}),
which requires the two \ensuremath{\HT{RecordChain}}s to
be ``pointwise equivalent'' meaning that the corresponding \ensuremath{\HSCon{Record}}s in the two \ensuremath{\HT{RecordChain}}s are equivalent.
A lemma \ensuremath{\HV{\hbox{\it RC\guydash{}irrelevant}}} shows that, if two record chains \ensuremath{\HSVar{rc}} and \ensuremath{\HV{rc^{\prime}}} end at the same \ensuremath{\HSCon{Record}}, then they
are equivalent (i.e., \ensuremath{\HSVar{rc}\;\mathrel{\HT{\approx\!\!\mathit{RC}}}\;\HV{rc^{\prime}}}), unless there is an injectivity failure.

The \ensuremath{\HT{\hbox{\it $\mathbb{K}$\guydash{}chain\guydash{}$\in\!$RC}}} property used in the proof of \ensuremath{\HSVar{propS4}} states that, if a \ensuremath{\HT{RecordChain}} \ensuremath{\HSVar{\mathit{rc}_{1}}} ends at a block \ensuremath{\HSVar{b}} that is
in a \ensuremath{\HT{\mathbb{K}\hbox{\it \guydash{}chain}}} based on another record chain \ensuremath{\HSVar{rc}}, then another \ensuremath{\HSCon{Block}} that is earlier in the \ensuremath{\HT{\mathbb{K}\hbox{\it \guydash{}chain}}} is also
``in'' \ensuremath{\HSVar{\mathit{rc}_{1}}}.  To enable proving this, \extended{the definition of} \ensuremath{\mathbin{\HT{\in\!\!RC}}} must allow for the possibility that the
other \ensuremath{\HSCon{Block}} is contained in an equivalent \ensuremath{\HT{RecordChain}}.  The definition of \ensuremath{\mathbin{\HT{\in\!\!RC}}} therefore has an additional
constructor beyond the two obvious ones, which enables the \ensuremath{\HSCon{Record}} in question to be
``transported'' from an equivalent \ensuremath{\HT{RecordChain}}:

\begin{myhs}
\begin{hscode}\SaveRestoreHook
\column{B}{@{}>{\hspre}l<{\hspost}@{}}%
\column{4}{@{}>{\hspre}l<{\hspost}@{}}%
\column{12}{@{}>{\hspre}l<{\hspost}@{}}%
\column{18}{@{}>{\hspre}l<{\hspost}@{}}%
\column{26}{@{}>{\hspre}l<{\hspost}@{}}%
\column{E}{@{}>{\hspre}l<{\hspost}@{}}%
\>[B]{}\HSKeyword{data}\;\HT{\_\!\in\!\!RC\_}\;{}\<[18]%
\>[18]{}\HSSpecial{\HSSym{\{\mskip1.5mu} }\HSVar{o}\;\mathbin{\HSCon{:}}\;\HSCon{Record}\HSSpecial{\HSSym{\mskip1.5mu\}}}\;\HSSpecial{(}\HSVar{\mathit{r}_{0}}\;\mathbin{\HSCon{:}}\;\HSCon{Record}\HSSpecial{)}\;\mathbin{\HSCon{:}}{}\<[E]%
\\
\>[18]{}\HSSym{\forall}\;\HSSpecial{\HSSym{\{\mskip1.5mu} }\HSVar{\mathit{r}_{1}}\HSSpecial{\HSSym{\mskip1.5mu\}}}\;\mathrel{\HSSym{\to}} \;\HT{RecordChainFrom}\;\HSVar{o}\;\HSVar{\mathit{r}_{1}}\;\mathrel{\HSSym{\to}} \;\HSCon{Set}\;\HSKeyword{where}{}\<[E]%
\\
\>[B]{}\hsindent{4}{}\<[4]%
\>[4]{}\HSVar{here}\;{}\<[12]%
\>[12]{}\mathbin{\HSCon{:}}\;\HSSym{\forall}\;\HSSpecial{\HSSym{\{\mskip1.5mu} }\HSVar{rc}\;\mathbin{\HSCon{:}}\;\HT{RecordChainFrom}\;\HSVar{o}\;\HSVar{\mathit{r}_{0}}\HSSpecial{\HSSym{\mskip1.5mu\}}}\;\mathrel{\HSSym{\to}} \;\HSVar{\mathit{r}_{0}}\;\mathbin{\HT{\in\!\!RC}}\;\HSVar{rc}{}\<[E]%
\\
\>[B]{}\hsindent{4}{}\<[4]%
\>[4]{}\HSVar{there}\;{}\<[12]%
\>[12]{}\mathbin{\HSCon{:}}\;\HSSym{\forall}\;\HSSpecial{\HSSym{\{\mskip1.5mu} }\HSVar{\mathit{r}_{1}}\;\HSVar{\mathit{r}_{2}}\HSSpecial{\HSSym{\mskip1.5mu\}}}\;\HSSpecial{\HSSym{\{\mskip1.5mu} }\HSVar{rc}\;\mathbin{\HSCon{:}}\;\HT{RecordChainFrom}\;\HSVar{o}\;\HSVar{\mathit{r}_{1}}\HSSpecial{\HSSym{\mskip1.5mu\}}}\;\mathrel{\HSSym{\to}} \;\HSSpecial{(}\HSVar{p}\;\mathbin{\HSCon{:}}\;\HSVar{\mathit{r}_{1}}\;\HT{\leftarrow}\;\HSVar{\mathit{r}_{2}}\HSSpecial{)}{}\<[E]%
\\
\>[12]{}\mathrel{\HSSym{\to}} \;\HSVar{\mathit{r}_{0}}\;\mathbin{\HT{\in\!\!RC}}\;\HSVar{rc}\;\mathrel{\HSSym{\to}} \;\HSVar{\mathit{r}_{0}}\;\mathbin{\HT{\in\!\!RC}}\;\HSSpecial{(}\HSVar{step}\;\HSVar{rc}\;\HSVar{p}\HSSpecial{)}{}\<[E]%
\\
\>[B]{}\hsindent{4}{}\<[4]%
\>[4]{}\HSVar{transp}\;{}\<[12]%
\>[12]{}\mathbin{\HSCon{:}}\;\HSSym{\forall}\;\HSSpecial{\HSSym{\{\mskip1.5mu} }\HSVar{r}\HSSpecial{\HSSym{\mskip1.5mu\}}}\;{}\<[26]%
\>[26]{}\HSSpecial{\HSSym{\{\mskip1.5mu} }\HSVar{rc0}\;\mathbin{\HSCon{:}}\;\HT{RecordChainFrom}\;\HSVar{o}\;\HSVar{r}\HSSpecial{\HSSym{\mskip1.5mu\}}}\;\HSSpecial{\HSSym{\{\mskip1.5mu} }\HSVar{\mathit{rc}_{1}}\;\mathbin{\HSCon{:}}\;\HT{RecordChainFrom}\;\HSVar{o}\;\HSVar{r}\HSSpecial{\HSSym{\mskip1.5mu\}}}{}\<[E]%
\\
\>[12]{}\mathrel{\HSSym{\to}} \;\HSVar{\mathit{r}_{0}}\;\mathbin{\HT{\in\!\!RC}}\;\HSVar{rc0}\;\mathrel{\HSSym{\to}} \;\HSVar{rc0}\;\mathrel{\HT{\approx\!\!\mathit{RC}}}\;\HSVar{\mathit{rc}_{1}}\;\mathrel{\HSSym{\to}} \;\HSVar{\mathit{r}_{0}}\;\mathbin{\HT{\in\!\!RC}}\;\HSVar{\mathit{rc}_{1}}{}\<[E]%
\ColumnHook
\end{hscode}\resethooks
\end{myhs}

\subsection{Traditional and extended correctness properties}
\label{sec:abstract:extended}

Our core correctness property \ensuremath{\HSCon{CommitsDoNotConflict}} is
\ensuremath{\HSVar{thmS5}} without the \ensuremath{\HT{\hbox{\it NonInjective\guydash{}$\equiv$}}} disjunct.  It
is proved in \ensuremath{\HSCon{LibraBFT.Abstract.Properties}}, which receives an
additional module parameter \ensuremath{\HT{\hbox{\it no\guydash{}collisions\guydash{}InSys}}} providing evidence that there are no injectivity failures between
\ensuremath{\HSCon{Block}}s that satisfy \ensuremath{\HSCon{InSys}}.  Note that, if an implementation reaches a state in which this does not hold,
then there is an injectivity failure between \emph{concrete} \ensuremath{\HSCon{Record}}s at the implementation level; for a typical
implementation, this signifies a collision for a cryptographic hash function among \ensuremath{\HSCon{Record}}s that are actually in
the system, contradicting the standard assumption that a computationally bounded adversary is unable to find such collisions.
To prove \ensuremath{\HSCon{CommitsDoNotConflict}}, we invoke \ensuremath{\HSVar{thmS5}} and
then use \ensuremath{\HT{\hbox{\it no\guydash{}collisions\guydash{}InSys}}} to eliminate the possibility of an injectivity failure.

To invoke \ensuremath{\HSCon{CommitsDoNotConflict}} for a particular implementation,
we need to provide \ensuremath{\HSCon{AllInSys}\;\HSVar{rc}}, where \ensuremath{\HSVar{rc}} is the \ensuremath{\HT{RecordChain}} for the first \ensuremath{\HSCon{CommitRule}} (and
similarly for \ensuremath{\HV{rc^{\prime}}}).  To enable this,
honest voters in typical implementations will vote to extend a \ensuremath{\HSCon{Block}}
only after verifying that the \ensuremath{\HSCon{Block}} extends a \ensuremath{\HSCon{QC}} (or the initial \ensuremath{\HSCon{Record}}) that the peer already knows
is in the system.
Thus, a peer that verifies a \ensuremath{\HSCon{CommitRule}} based on a record chain \ensuremath{\HSVar{rc}}
that ends in a \ensuremath{\HSCon{QC}} (\ensuremath{\HSVar{q}}) knows that every \ensuremath{\HSCon{Record}} in \ensuremath{\HSVar{rc}} is ``in the system'':
\ensuremath{\HSCon{AllInSys}\;\HSVar{rc}}.

\paragraph{Extended correctness condition}  We are also interested in enabling parties
that do not participate in the protocol to verify commits.
Suppose a peer \ensuremath{\HSVar{p}} provides to a client \ensuremath{\HSVar{c}} the contents of a \ensuremath{\HSCon{CommitRule}} that
\ensuremath{\HSVar{c}} can verify.  In this case, \ensuremath{\HSVar{c}} cannot invoke \ensuremath{\HSCon{CommitsDoNotConflict}}
(or \ensuremath{\HSVar{thmS5}}), because it does not know the \ensuremath{\HT{RecordChain}} on which the \ensuremath{\HSCon{CommitRule}} is based.

For this purpose,
we define and prove a variant of \ensuremath{\HSCon{CommitsDoNotConflict}} called
\ensuremath{\HT{\hbox{\it CommitsDoNotConflict$^{\prime}$}}}.  This condition ensures that
even a party that does not participate in consensus
can confirm commits and will not confirm conflicting commits.

\begin{myhs}
\begin{hscode}\SaveRestoreHook
\column{B}{@{}>{\hspre}l<{\hspost}@{}}%
\column{3}{@{}>{\hspre}l<{\hspost}@{}}%
\column{12}{@{}>{\hspre}l<{\hspost}@{}}%
\column{14}{@{}>{\hspre}l<{\hspost}@{}}%
\column{49}{@{}>{\hspre}l<{\hspost}@{}}%
\column{E}{@{}>{\hspre}l<{\hspost}@{}}%
\>[B]{}\HT{\hbox{\it CommitsDoNotConflict$^{\prime}$}}\;\mathbin{\HSCon{:}}\;\HSSym{\forall}\;\HSSpecial{\HSSym{\{\mskip1.5mu} }\HSVar{o}\;\HV{o^{\prime}}\;\HSVar{q}\;\HV{q^{\prime}}\HSSpecial{\HSSym{\mskip1.5mu\}}}{}\<[E]%
\\
\>[B]{}\hsindent{3}{}\<[3]%
\>[3]{}\mathrel{\HSSym{\to}} \;\HSSpecial{\HSSym{\{\mskip1.5mu} }\HSVar{rcf}\;{}\<[12]%
\>[12]{}\mathbin{\HSCon{:}}\;\HT{RecordChainFrom}\;\HSVar{o}\;\HSSpecial{(}\HSCon{Q}\;\HSVar{q}\HSSpecial{)}\HSSpecial{\HSSym{\mskip1.5mu\}}}\;{}\<[49]%
\>[49]{}\mathrel{\HSSym{\to}} \;\HSCon{AllInSys}\;\HSVar{rcf}{}\<[E]%
\\
\>[B]{}\hsindent{3}{}\<[3]%
\>[3]{}\mathrel{\HSSym{\to}} \;\HSSpecial{\HSSym{\{\mskip1.5mu} }\HV{rc^{\prime}}\;\mathbin{\HSCon{:}}\;\HT{RecordChainFrom}\;\HSVar{o'}\;\HSSpecial{(}\HSCon{Q}\;\HV{q^{\prime}}\HSSpecial{)}\HSSpecial{\HSSym{\mskip1.5mu\}}}\;{}\<[49]%
\>[49]{}\mathrel{\HSSym{\to}} \;\HSCon{AllInSys}\;\HV{rc^{\prime}}{}\<[E]%
\\
\>[B]{}\hsindent{3}{}\<[3]%
\>[3]{}\mathrel{\HSSym{\to}} \;\HSSpecial{\HSSym{\{\mskip1.5mu} }\HSVar{b}\;\HV{b^{\prime}}\;\mathbin{\HSCon{:}}\;\HSCon{Block}\HSSpecial{\HSSym{\mskip1.5mu\}}}\;\mathrel{\HSSym{\to}} \;\HSCon{CommitRuleFrom}\;\HSVar{rcf}\;\HSVar{b}\;\mathrel{\HSSym{\to}} \;\HSCon{CommitRuleFrom}\;\HV{\hbox{\it rcf$^{\prime}$}}\;\HV{b^{\prime}}{}\<[E]%
\\
\>[B]{}\hsindent{3}{}\<[3]%
\>[3]{}\mathrel{\HSSym{\to}} \;\HSCon{Either}\;{}\<[14]%
\>[14]{}\HTNI{\Sigma}\;\HSSpecial{(}\HT{RecordChain}\;\HSSpecial{(}\HSCon{Q}\;\HV{q^{\prime}}\HSSpecial{)}\HSSpecial{)}\;\HSSpecial{(}\HSSpecial{(}\HSCon{B}\;\HSVar{b}\HSSpecial{)}\;\mathbin{\HT{\in\!\!RC}\_}\HSSpecial{)}\;{}\<[E]%
\\
\>[14]{}\HTNI{\Sigma}\;\HSSpecial{(}\HT{RecordChain}\;\HSSpecial{(}\HSCon{Q}\;\HSVar{q}\HSSpecial{)}\HSSpecial{)}\;\HSSpecial{(}\HSSpecial{(}\HSCon{B}\;\HV{b^{\prime}}\HSSpecial{)}\;\mathbin{\HT{\in\!\!RC}\_}\HSSpecial{)}{}\<[E]%
\ColumnHook
\end{hscode}\resethooks
\end{myhs}

\ensuremath{\HT{\hbox{\it CommitsDoNotConflict$^{\prime}$}}} does not require \ensuremath{\HSCon{CommitRule}}s based on full \linebreak
\ensuremath{\HT{RecordChain}}s; instead, \ensuremath{\HSCon{CommitRuleFrom}}s based on \ensuremath{\HT{RecordChainFrom}}s suffice.  This property shows that
a party can validate just the \ensuremath{\HSCon{Record}}s required to form a \ensuremath{\HSCon{CommitRuleFrom}}, and confirm
that the \ensuremath{\HSCon{Block}} it claims to commit has indeed been committed, and that there cannot be another
commit that conflicts with it.  Here, \ensuremath{\HSSpecial{(}\HSCon{B}\;\HSVar{b}\HSSpecial{)}\;\mathbin{\HT{\in\!\!RC}\_}}
is a predicate over values of type \ensuremath{\HT{RecordChain}\;\HSSpecial{(}\HSCon{Q}\;\HV{q^{\prime}}\HSSpecial{)}}, so \ensuremath{\HT{\hbox{\it CommitsDoNotConflict$^{\prime}$}}} says
that, if there are two \ensuremath{\HSCon{CommitRuleFrom}}s based on \ensuremath{\HT{RecordChainFrom}}s that end with a \ensuremath{\HSCon{QC}} and have all
of their \ensuremath{\HSCon{Record}}s in the system, then (unless there is an injectivity failure), one of committed \ensuremath{\HSCon{Block}}s
is in a \ensuremath{\HT{RecordChain}} that contains the other.

To prove this property, we require an additional assumption about the implementation, which is
provided as a module parameter \ensuremath{\mathrel{\HT{\in\!\mathit{\hbox{QC\ensuremath{\Rightarrow{}}\!\!AllSent}}}}}, of type \ensuremath{\HSCon{Complete}\;\HSCon{InSys}}, where:

\begin{myhs}
\begin{hscode}\SaveRestoreHook
\column{B}{@{}>{\hspre}l<{\hspost}@{}}%
\column{17}{@{}>{\hspre}l<{\hspost}@{}}%
\column{30}{@{}>{\hspre}l<{\hspost}@{}}%
\column{E}{@{}>{\hspre}l<{\hspost}@{}}%
\>[B]{}\HSCon{Complete}\;\mathbin{\HSCon{:}}\;\HSSym{\forall}\;\HSSpecial{\HSSym{\{\mskip1.5mu} }\HVNI{\ell}\HSSpecial{\HSSym{\mskip1.5mu\}}}\;\mathrel{\HSSym{\to}} \;\HSSpecial{(}\HSCon{Record}\;\mathrel{\HSSym{\to}} \;\HSCon{Set}\;\HVNI{\ell}\HSSpecial{)}\;\mathrel{\HSSym{\to}} \;\HSCon{Set}\;\HVNI{\ell}{}\<[E]%
\\
\>[B]{}\HSCon{Complete}\;\mathbin{\HT{\in\!\!sys}}\;{}\<[17]%
\>[17]{}\mathrel{\HSSym{=}}\;\HSSym{\forall}\;\HSSpecial{\HSSym{\{\mskip1.5mu} }\HSVar{a}\;\HSVar{q}\HSSpecial{\HSSym{\mskip1.5mu\}}}\;\mathrel{\HSSym{\to}} \;\HSCon{MetaHonestMember}\;\HSVar{a}{}\<[E]%
\\
\>[17]{}\mathrel{\HSSym{\to}} \;\HSVar{a}\;\mathbin{\HT{\in\!\!QC}}\;\HSVar{q}\;\mathrel{\HSSym{\to}} \;\mathbin{\HT{\in\!\!sys}}\;\HSSpecial{(}\HSCon{Q}\;\HSVar{q}\HSSpecial{)}{}\<[E]%
\\
\>[17]{}\mathrel{\HSSym{\to}} \;\HT{\exists[}{\HSVar{b}}\HT{]}\;{}\<[30]%
\>[30]{}\HSSpecial{(}\HTNI{\Sigma}\;\HSSpecial{(}\HT{RecordChain}\;\HSSpecial{(}\HSCon{B}\;\HSVar{b}\HSSpecial{)}\HSSpecial{)}\;\HSCon{AllInSys}\;\mathbin{\HT{\times}}\;\HSCon{B}\;\HSVar{b}\;\HT{\leftarrow}\;\HSCon{Q}\;\HSVar{q}\HSSpecial{)}{}\<[E]%
\ColumnHook
\end{hscode}\resethooks
\end{myhs}

Here, \ensuremath{\HSCon{Record}\;\mathrel{\HSSym{\to}} \;\HSCon{Set}\;\HVNI{\ell}} is a predicate on (abstract) \ensuremath{\HSCon{Record}}s representing what
\ensuremath{\HSCon{Record}}s an implementation considers to be ``in the system''.

This assumption (indirectly) requires that an honest peer sends a \ensuremath{\HSCon{Vote}} for a \ensuremath{\HSCon{Block}} id (which may
subsequently be represented in a \ensuremath{\HSCon{QC}}) only if it knows that there is a \ensuremath{\HSCon{Block}} with that id
and a \ensuremath{\HT{RecordChain}} up to that \ensuremath{\HSCon{Block}}
whose \ensuremath{\HSCon{Record}}s are all ``in the system'' (for example the peer may have validated all of those \ensuremath{\HSCon{Record}}s itself,
or it may have validated sufficient information to be confident that all of them have been validated by some
honest peer, unless there is a hash collision among \ensuremath{\HSCon{Record}}s that are in the system).

\extendedalt{Next, we overview the proof
for \ensuremath{\HT{\hbox{\it CommitsDoNotConflict$^{\prime}$}}}, and for a lemma \ensuremath{\HV{\mathit{\hbox{crf\ensuremath{\Rightarrow{}}\!\!cr}}}} on which it depends.

\paragraph{\bf Overview of proof for \ensuremath{\HT{\hbox{\it CommitsDoNotConflict$^{\prime}$}}}:}
The proof of this property can be found in \ensuremath{\HSCon{LibraBFT.Abstract.Properties}}.
It constructs arguments sufficient to invoke \ensuremath{\HSCon{CommitsDoNotConflict}} (aka \ensuremath{\HSVar{thmS5}})
and uses its return value to construct the necessary result.

It first uses the \ensuremath{\HV{\hbox{\it bft\guydash{}assumption}}} provided by the \ensuremath{\HSCon{EpochConfig}} to determine an honest peer represented
in both \ensuremath{\HSCon{QC}}s (\ensuremath{\HSCon{Q}\;\HSVar{q}} and \ensuremath{\HSCon{Q}\;\HV{q^{\prime}}}), and then invokes the \ensuremath{\mathrel{\HT{\in\!\mathit{\hbox{QC\ensuremath{\Rightarrow{}}\!\!AllSent}}}}} module parameter
to obtain a \ensuremath{\HSCon{Block}} \ensuremath{\HSVar{b}} and a \ensuremath{\HT{RecordChain}} up to \ensuremath{\HSCon{B}\;\HSVar{b}} whose \ensuremath{\HSCon{Record}}s are all \ensuremath{\HSCon{InSys}}, as well as evidence
that \ensuremath{\HSVar{q}} extends \ensuremath{\HSVar{b}}.  It similarly uses \ensuremath{\mathrel{\HT{\in\!\mathit{\hbox{QC\ensuremath{\Rightarrow{}}\!\!AllSent}}}}} to obtain analogous information for \ensuremath{\HV{q^{\prime}}}.

We are not yet ready to invoke \ensuremath{\HSCon{CommitsDoNotConflict}}, however, because
there is no guarantee that the \ensuremath{\HT{RecordChain}}s acquired using \ensuremath{\mathrel{\HT{\in\!\mathit{\hbox{QC\ensuremath{\Rightarrow{}}\!\!AllSent}}}}} are identical to the ones on which the relevant
\ensuremath{\HSCon{CommitRuleFrom}}s are based.

To solve this, we invoke a lemma \ensuremath{\HV{\mathit{\hbox{crf\ensuremath{\Rightarrow{}}\!\!cr}}}}, which proves that if we have a \ensuremath{\HT{RecordChain}\;\HSSpecial{(}\HSCon{Q}\;\HSVar{q}\HSSpecial{)}}
and a \ensuremath{\HT{RecordChainFrom}\;\HSSpecial{(}\HSCon{Q}\;\HSVar{q}\HSSpecial{)}} (call it \ensuremath{\HSVar{rcf}}), and a \linebreak \ensuremath{\HSCon{CommitRuleFrom}\;\HSVar{rcf}\;\HSVar{b}} for some block \ensuremath{\HSVar{b}}, then we can construct a \linebreak
\ensuremath{\HSCon{CommitRule}\;\HSVar{rc}\;\HSVar{b}} (unless there is an injectivity failure).\qed

\paragraph{\bf Overview of proof for \ensuremath{\HV{\mathit{\hbox{crf\ensuremath{\Rightarrow{}}\!\!cr}}}}:}
The proof of this property is also in \linebreak \ensuremath{\HSCon{LibraBFT.Abstract.RecordChain}}.
It uses properties about \ensuremath{\HT{RecordChainFrom}} like those
for \ensuremath{\HT{RecordChain}}.

A lemma \ensuremath{\HV{\hbox{\it RCF\guydash{}irrelevant}}} shows that, given two
\ensuremath{\HT{RecordChainFrom}}s that end at the same \ensuremath{\HSCon{Record}}, one is pointwise equivalent to a suffix of the other
(as usual, modulo injectivity failures).

For the case in which the \ensuremath{\HT{RecordChainFrom}} is a suffix of the \ensuremath{\HT{RecordChain}}, we invoke
a lemma \ensuremath{\HV{\hbox{\it transp\guydash{}k\guydash{}chain\guydash{}$\subseteq$}}} to ``transport'' the \ensuremath{\HSVar{kchain}}
on which the \ensuremath{\HSCon{CommitRuleFrom}} is based to a \ensuremath{\HSCon{CommitRule}} based on the \ensuremath{\HT{RecordChain}}, and another
lemma \ensuremath{\HV{\hbox{\it kchainBlock\guydash{}$\subseteq$RC}}}, which shows that this transporting does not change the committed \ensuremath{\HSCon{Block}}.
We then use the results of these invocations
to construct a \ensuremath{\HSCon{CommitRule}} based on the \ensuremath{\HT{RecordChain}}, as required to invoke \linebreak \ensuremath{\HSCon{CommitsDoNotConflict}}.

In the other case, when the \ensuremath{\HT{RecordChain}}
is a suffix of the \ensuremath{\HT{RecordChainFrom}}, we use some simple properties to conclude that the \ensuremath{\HT{RecordChainFrom}}
must also be a \ensuremath{\HT{RecordChain}}, and then use \ensuremath{\HV{\hbox{\it transp\guydash{}CR}}} to transport the original \linebreak \ensuremath{\HSCon{CommitRuleFrom}} (which is
in fact a \ensuremath{\HSCon{CommitRule}} in this case) to the original \ensuremath{\HT{RecordChain}}.\qed
}{The extended version of this paper~\cite{librabft-agda-extended} includes proof overviews
for \linebreak \ensuremath{\HT{\hbox{\it CommitsDoNotConflict$^{\prime}$}}}, and for a lemma \ensuremath{\HV{\mathit{\hbox{crf\ensuremath{\Rightarrow{}}\!\!cr}}}} on which it depends.}

\subsection{Relating non-injectivity to security properties}
\label{sec:hashinj}

Recall from \Cref{sec:thmsfive} that we prove our abstract properties modulo injectivity of
\ensuremath{\HSCon{Block}} ids.  However, the simplified \ensuremath{\HT{\hbox{\it NonInjective\guydash{}$\equiv$}}} disjunct used in the property definitions
presented so far is insufficient.  The reason is that it is \emph{trivial} to construct two different
abstract \ensuremath{\HSCon{Block}}s with the same id, meaning that we could prove \ensuremath{\HSVar{thmS5}} with a single-line proof,
independent of the actual protocol.  Worse, we could accidentally do the same in context of
legitimate-looking proofs.

The issue is that the abstract \ensuremath{\HSCon{Block}}s we could trivially construct bear no relation to any real
\ensuremath{\HSCon{Block}}s and ids produced in the execution of a concrete implementation.  To resolve this problem,
we strengthen the first disjunct of \ensuremath{\HSVar{thmS5}} to \ensuremath{\HT{\hbox{\it NonInjective\guydash{}$\equiv$\guydash{}InSys}}}, defined as follows:

\begin{myhs}
\begin{hscode}\SaveRestoreHook
\column{B}{@{}>{\hspre}l<{\hspost}@{}}%
\column{3}{@{}>{\hspre}l<{\hspost}@{}}%
\column{E}{@{}>{\hspre}l<{\hspost}@{}}%
\>[B]{}\HT{\hbox{\it NonInjective\guydash{}$\equiv$\guydash{}InSys}}\;\mathbin{\HSCon{:}}\;\HSCon{Set}{}\<[E]%
\\
\>[B]{}\HT{\hbox{\it NonInjective\guydash{}$\equiv$\guydash{}InSys}}\;\mathrel{\HSSym{=}}\;{}\<[E]%
\\
\>[B]{}\hsindent{3}{}\<[3]%
\>[3]{}\HTNI{\Sigma}\;\HT{\hbox{\it NonInjective\guydash{}$\equiv$}}\;\HSSym{\lambda} \;\HSSpecial{\HSSym{\{\mskip1.5mu} }\HSSpecial{(}\HSSpecial{(}\HSVar{\mathit{b}_{0}}\;\mathbin{\HSSym{,}}\;\HSVar{\mathit{b}_{1}}\HSSpecial{)}\;\mathbin{\HSSym{,}}\;\mathbin{\HSSym{\anonymous}} \;\mathbin{\HSSym{,}}\;\mathbin{\HSSym{\anonymous}} \HSSpecial{)}\;\mathrel{\HSSym{\to}} \;\HSCon{InSys}\;\HSSpecial{(}\HSCon{B}\;\HSVar{\mathit{b}_{0}}\HSSpecial{)}\;\mathbin{\HT{\times}}\;\HSCon{InSys}\;\HSSpecial{(}\HSCon{B}\;\HSVar{\mathit{b}_{1}}\HSSpecial{)}\HSSpecial{\HSSym{\mskip1.5mu\}}}{}\<[E]%
\ColumnHook
\end{hscode}\resethooks
\end{myhs}

This definition requires that the proof not only provides different \ensuremath{\HSCon{Block}}s \ensuremath{\HSVar{\mathit{b}_{0}}} and \ensuremath{\HSVar{\mathit{b}_{1}}} with the
same id, but also proof that the implementation considers the \ensuremath{\HSCon{Record}}s \ensuremath{\HSCon{B}\;\HSVar{\mathit{b}_{0}}} and \ensuremath{\HSCon{B}\;\HSVar{\mathit{b}_{1}}} to be ``in
the system''\!.  The meaning of ``in the system'' is specified by the implementation-provided
predicate \ensuremath{\HSCon{InSys}} and is thus beyond the scope of this paper.  However, in ongoing work, we are
proving a real implementation correct using the results presented here.  In that broader context, we
instantiate \ensuremath{\HSCon{InSys}} with a predicate that holds only for \ensuremath{\HSCon{Block}}s that are contained in network
messages that have actually been sent.  In this way, from the perspective of that concrete
implementation, we ensure that our correctness properties hold unless and until an adversary
\emph{actually finds a hash collision} and introduces it into the system.  We contrast this approach
to some related efforts in~\Cref{sec:relwork}.

The \ensuremath{\HT{\hbox{\it NonInjective\guydash{}$\equiv$}}} and \ensuremath{\HT{\hbox{\it NonInjective\guydash{}$\equiv$\guydash{}InSys}}} definitions stated above are actually simplified
versions of more general definitions we use in our proofs; details are available in our open source
development~\cite{librabft-agda}.  These more general definitions are required because, at different
stages of our proofs, we use different predicates to capture evidence collected so far about the
conflicting \ensuremath{\HSCon{Block}}s, so that we can build up to the proof for \ensuremath{\HSVar{thmS5}} that both \ensuremath{\HSCon{Block}}s satisfy
\ensuremath{\HSCon{InSys}}.

\section{Related work}
\label{sec:relwork}

\subsection{\hotstuff{}/\librabft{}}

Before open sourcing our work in December 2020~\cite{librabft-agda}, we were
not aware of any formal verification work related to the \hotstuff{} / \librabft{} protocols beyond
manual proof sketches~\cite{hotstuff-podc-19, libra-2019-06-28, libra-2020-05-26};
these are useful and have influenced our work significantly, but are far from detailed,
precise proofs.  We have since learned of two other pieces of work involving mechanical proofs
of correctness of variants of the \hotstuff{}/\librabft{} algorithm, and one involving model checking.

Librachain~\cite{librachain} is a Coq-based model of the data structures used in \librabftorig{}.  It contains a single
commit from May 2020, described as ``experimental''; we are not aware of any paper
describing this work.  The Librachain model commits
to some structural details that are not central to the core protocol\extended{ underlying \hotstuff{} and \librabft{}}.
For example, it assumes that the \ensuremath{\HSCon{QuorumCert}} that a new
\ensuremath{\HSCon{Block}} extends is included in the \ensuremath{\HSCon{Block}} record; this is one implementation choice, but certainly
not fundamental.  Furthermore, the proofs assume various conditions have been validated for the
data structures, and are thus intimately tied to the particular implementation types.
In contrast, we model an \emph{abstraction} of the core protocol, and establish
precise requirements for \emph{any} implementation to enjoy the correctness properties we prove.
\extended{Thus, we enable proving correctness for a variety of practical
implementations by proving only that they satisfy these requirements.}  The Librachain development
also uses a hypothesis that the hash function used is injective, which is not true of hash functions that
are used in practice.  Our properties are proved to hold unless and until a \emph{specific} injectivity failure
exists between (abstract) \ensuremath{\HSCon{Record}}s that are actually ``in the system''
(see~\Cref{sec:hashinj}); when instantiated with implementations that use
cryptographic hash functions to assign ids, this ensures that the result holds unless and until a peer succeeds
in finding a specific hash collision, violating the assumption that a computationally bounded adversary cannot do so.

More recently, Leander~\cite{DBLP:conf/forte/Jehl21} has described work modeling and proving
correctness for one specific, simplified variant of \hotstuff{}.  Hashes are not explicitly modeled,
but the way the relationship between blocks is modeled amounts to an assumption that hashing is
injective.  Leander modeled this simplified variant in TLA+ and Ivy, and the paper is focused on
comparing the tools for this purpose.

Kukharenko et al.~\cite{hotstuff-tla} use TLA+~\cite{Kuppe_2019} to model check
\emph{basic} \hotstuff{}, but not the more practical
\emph{chained} variant used by \librabft{}.
Again, our work applies to an abstraction of the
protocol that can be instantiated for all versions of \hotstuff{} and \librabft{}, as well
as variants that may not yet exist.

Model checking has the advantage of requiring less work (defining a model and correctness
properties and then ``pushing the button'') than developing precise, machine-checked
correctness proofs.
It can also provide insight into errors found.  Kukharenko et al. ran one of
their models with seven participants of which three are byzantine
(correctness is not guaranteed in this case), and found a counterexample showing \emph{how} the
byzantine peers can violate correctness.

\extended{However, there are also some drawbacks to the model checking approach.}
To limit the state space, Kukharenko
et al. developed a restricted model, in which a node (analogous to our \ensuremath{\HSCon{Block}}) can be extended only by
one of two nodes, and a more general model in which any node can extend any other (from some fixed set).
The restricted model, \extended{configured} with just four peers (one \extended{of which is} byzantine), took over seven hours to check.
The more general model took over 17 days.  Our approach imposes no such limitations, and Agda checks our proofs
in under one minute.
Finally, for the more general model, TLA+ estimates an ``optimistic'' probability of 0.3 that it has in
fact not explored the entire state space due to hash collisions on states, leaving open the possibility of
an unfound bug even for this minimal configuration.
We consider that Kukharenko et al.'s work complements ours,
but does not obviate the need for the machine-checked correctness proofs.

\subsection{Other BFT consensus protocols}

P\^{\i}rlea and Sergey present Toychain~\cite{10.1145/3167086,toychain-thesis}, which models
Nakamoto consensus~\cite{nakamoto} and proves correctness properties about it using Coq~\cite{10.5555/1965123}.  Although Nakamoto
consensus differs substantially from \hotstuff{}/\librabft{}\extended{, including that it is not actually
consensus in the traditional sense, because agreement is not final}, Toychain is the
closest prior work to ours in terms of modeling structures (collections of trees of records) and reasoning about their properties.
\extendedalt{Specifically, t}{T}heir model can be instantiated with different
implementation components, and they prove that any implementation that
provides components satisfying certain requirements is correct.
In contrast, each of the \librabft{}-related efforts mentioned above \cite{librachain, DBLP:conf/forte/Jehl21,
hotstuff-tla} proves properties about one particular model of \hotstuff{}/\librabft{}.

While Toychain indeed establishes some generality by enabling instantiation with specific
components, we impose no structure whatsoever on an implementation: if the externally visible
behaviour of honest peers for a given implementation complies with two precisely stated rules, then
that implementation can inherit the correctness properties we have proved of the abstract model.

Toychain initially assumed an injective hash function\extendedalt{.  Because real hash
functions are not injective, this would require}{, which requires} trusting that the proofs do not abuse the power
granted by \extended{depending on} a false assumption.  Interestingly, subsequent versions of Toychain
addressed this issue by removing \extendedalt{some unsatisfiable assumptions, including}{the assumption} that the hash function
used is injective.  The bulk of Chapter 3 of P\^{\i}rlea's
thesis~\cite{toychain-thesis} is devoted to describing the complexity that this undertaking
involved, reporting that \emph{every} proof had to be changed, and citing an example of one proof
that grew from 10 lines to 150 to accommodate this enhancement!

In contrast, as described in~\Cref{sec:hashinj}, we have taken a different approach.
Our abstract
model is aware only of ids assigned to \ensuremath{\HSCon{Block}}s that an implementation
\extended{instantiating our model} considers to be ``in the system''\!, not hash functions.  \extendedalt{Like P\^{\i}rlea and Sergey, we}{We}
too rested our initial development on an unsound
foundation by assuming that ids were injective.  However, because our abstraction freed us from reasoning about hash functions in our
correctness proofs, it was not particualrly disruptive to later augment our proofs to provide
evidence of \emph{specific} injectivity failures when necessary,
tying those injectivity failures to \ensuremath{\HSCon{Record}}s that the implementation considers to be in the
system.    \extended{An implementation can then use our results to establish correctness
properties that hold unless and until a peer \emph{actually finds a specific collision} for the particular hash function
used by the implementation. Recall that it is a standard assumption that a computationally bounded
adversary cannot do so~\citep[Chapter 9]{10.5555/548089}.}

The work that is perhaps closest to our broader project is
Velisarios~\cite{10.1007/978-3-319-89884-1_22}, which uses the Coq theorem
prover~\cite{10.5555/1965123} and provides a framework for modeling distributed systems with
byzantine peers, analogous to our system model.  It is based on a Logic of
Events~\cite{BCR12} approach, in contrast to our state transition system approach.  Velisarios
is instantiated with definitions modeling PBFT~\cite{10.5555/296806.296824}
to prove PBFT correct.
Coq supports extraction to OCaml, enabling an implementation to be derived from the PBFT model.
Agda has support for extracting to Haskell or Javascript.  However, we have not experimented with this.
The goal of our ongoing work is to model our practical Haskell implementation in Agda and prove correctness for
that model using the results presented in this paper.  \extended{Asphalion~\cite{10.1145/3360564} takes the direction of Velisarios further, enabling reasoning about
\emph{hybrid} system models in which different components require different failure assumptions, for example
to model the use of trusted computing components.}

Alturki et al.~\cite{Alturki2020} use Coq to formally verify correctness of
Algorand's~\cite{10.1145/3132747.3132757} consensus protocol.  Their correctness condition is slightly
different as Algorand's protocol seeks to ensure that exactly one block is certified per round, implying
a total order on all certified blocks.
Crary~\cite{crary2021verifying} reports on work towards verifying correctness
for the consensus mechanism of Hashgraph~\cite{hashgraph} in Coq.
Losa and Dodds~\cite{stellar-fmbc-2020} describe formal verification of safety and liveness properties
for the Stellar consensus protocol using Ivy and Isabelle/HOL.
Alturki et al.~\cite{gasper-coq} use Coq to formally verify properties for
Gasper~\cite{buterin2020combining}---Ethereum 2.0's Proof of Stake
consensus mechanism.  Rather than assuming that
any two quorums intersect on at least one honest node, they prove that, if (using our
terminology) two conflicting blocks are committed, then there exist two quorums whose
common members can have their stake slashed.  This property
would be satisfied if only the first offense results in slashing; presumably,
a stronger property that ties the conflicting blocks to specific quorums related to \extended{committing}those
blocks could be proved.

There is also work model checking other BFT consensus protocols.  For example,
Tholoniat and Gramoli~\cite{redbelly-2019} have used ByMC~\cite{DBLP:conf/isola/0001W18}
to model check RedBelly's consensus algorithm~\cite{8548057}; ByMC is a model checker designed
to mitigate the state space blowup for algorithms in which processes wait
for a threshold of messages. While basic \hotstuff{} may fit this structure, chained \hotstuff{}
does not.  \extended{Furthermore, \librabft{} defines quorums in terms of combined voting power, not
just the number of peers voting.}

Braithwaite et al.~\cite{tendermint-model-check-fmbc-2020} report on work in progress towards model
checking Tendermint~\cite{buchman2019latest} using TLA+; so far, they have gained
useful insight into the algorithm using very small configurations, and have found and fixed some
specification bugs as a result.  Nonetheless, their experience again highlights
\extendedalt{the challenges
and limitations of the model checking approach due to the need to limit the state space size to
achieve acceptable execution time.}{the challenges of model checking related to state space and execution time.}

\section{Concluding remarks and future work}
\label{sec:conc}

We have presented a formal model of the essence of a Byzantine Fault Tolerant consensus protocol
used in several existing implementations, and proved its safety properties---including
one that enables non-participants to verify commits---for a single epoch, during which configuration
does not change.  Extending our proofs to accommodate epoch changes (\extended{also known as }reconfiguration)
is future work.

Our contributions include precisely defining implementation assumptions
and correctness conditions,
and developing formal, machine-checked proofs of correctness properties for any implementation satisfying the
assumptions.
Our model, definitions, and proofs are all
expressed in Agda, and are available in open source.

Our approach enables verifying implementations by proving only that
honest peers obey the rules established by our abstract assumptions, without repeating the
hard work of proving the underlying protocol correct each time.

Our \ensuremath{\HSVar{thmS5}} property establishes correctness unless it can provide \emph{evidence} of a
\emph{specific} injectivity failure between \ensuremath{\HSCon{Block}}s that are \emph{in the system}.  Thus our
proofs are independent of how specific implementations assign \ensuremath{\HSCon{Block}} ids, and
ensure that they hold unless and until an injectivity failure actually occurs.  In this way, our
abstract proofs support proving that implementations that use crypotgraphic hash
functions to assign ids behave correctly, based on the standard assumption that a computationally bounded
adversary cannot produce a hash collision.

In our broader project~\cite{librabft-agda}, we have
defined a system model \extendedalt{that can be instantiated \extended{with types and handlers} for a particular
implementation\extended{, resulting in a model of a system operating with that implementation}.}{in which}
messages can be lost, duplicated and arbitrarily delayed, and dishonest peers are
constrained only by their inability to forge signatures of honest peers.
We have ported our Haskell implementation to Agda using a library we have developed~\cite{nasa-submission-syntax-short},
instantiated our system model with its types and handlers,
and made substantial progress towards proving that it satisfies the required assumptions.

Beyond that,
extending our system model to support proofs of liveness in the partial synchrony model~\cite{10.1145/42282.42283}
is future work.  A pragmatic intermediate point is to prove within our existing system model\extended{ a temporal logic property
stating} that, from any reachable state that has \ensuremath{\HSCon{Block}}s available to commit, there is some execution
in which another \ensuremath{\HSCon{Block}} is committed (called \emph{plausible liveness} by Buterin and Griffith~\cite{DBLP:journals/corr/abs-1710-09437}).
These liveness properties would pertain to a
model of a specific \emph{implementation}; liveness properties do not make sense for the abstract model
presented in this paper.
\lhsinclude{body.lhs}

\bibliography{references}
\end{document}